\documentclass{aa}

\usepackage{amsmath}
\usepackage{bbold}
\usepackage{natbib}
\usepackage{graphicx}
\usepackage{tabularx}
\usepackage[percent]{overpic}
\usepackage{xcolor}

\newcommand{\corr}[1]{#1}

\bibpunct{(}{)}{;}{a}{}{,}

\begin{document}
 
\title{Tomography of the Galactic free electron density with the Square Kilometer Array}

\author{M.~Greiner\thanks{\email{maksim@mpa-garching.mpg.de}}\inst{\ref{inst:MPA}}
  \and D.H.F.M.~Schnitzeler\inst{\ref{inst:MPIFR}}
  \and T.A.~En\ss{}lin\inst{\ref{inst:MPA}}
  }

\institute{Max-Planck-Institut f\"ur Astrophysik, Karl-Schwarzschild-Str.~1, 85748 Garching, Germany \label{inst:MPA}
  \and Max-Planck-Institut f\"ur Radioastronomie, Auf dem H\"ugel 69, 53121 Bonn, Germany \label{inst:MPIFR}}

\date{Received DD MMM. YYYY / Accepted DD MMM. YYYY}

\abstract{
We present a new algorithm to reconstruct the Galactic free electron density from pulsar dispersion measures. The algorithm performs a nonparametric tomography for a density field with an arbitrary amount of degrees of freedom. It is based on approximating the Galactic free electron density as the product of a profile function with a statistically isotropic and homogeneous log-normal field. Under this approximation the algorithm generates a map of the free electron density as well as an uncertainty estimate without the need of information about the power spectrum. The uncertainties of the pulsar distances are treated consistently by an iterative procedure.
We test the algorithm using the NE2001 model with modified fluctuations as a Galaxy model, pulsar populations generated from the Lorimer population model, and mock observations emulating the upcoming Square Kilometer Array.
We show the quality of the reconstruction for mock data sets containing between 1000 and 10000 pulsars with distance uncertainties up to 25\%. Our results show, that with the SKA nonparametric tomography of the Galactic free electron density becomes feasible, but the quality of the reconstruction is very sensitive to the distance uncertainties.
}


\titlerunning{Reconstructing the Galactic free electron density}
\authorrunning{M.~Greiner et al.}
\maketitle

\section{Introduction}

Regions of the interstellar medium that are (partly) ionized play an important role in a number of effects
\corr{such as pulse dispersion and scattering, and Faraday rotation. }
Additionally, ionized parts of the interstellar medium emit radiation through free-free emission and $\mathrm{H}_\alpha$ emission. The magnitude of these effects depends on the distribution of free electrons, the \textit{free electron density}. It is therefore of great interest to model or reconstruct the free electron density as accurately as possible. 

Reconstruction and modeling of the Milky Way has been an ongoing topic of research for many years. The free electron density has been modeled by \cite{Taylor-1993}, \cite{cordes-2002}, and \cite{Gaensler-2008} among others. For a comparison and discussion of various existing models see \cite{Schnitzeler-2012} and for a review of the mapping of HI regions see \cite{Kalberla-2009}. The interstellar magnetic \corr{field} has been modeled by \cite{Sun-2008,Sun-2010} and \cite{Jansson-2012,Jansson-2012b}. The dust distribution has been modeled by e.g.~\cite{Berry-2012} and even \corr{nonparametric} tomography has been performed by \cite{Lallement-2014} and \cite{Sale-2014}.

\corr{We plan to use the dispersion measures ($D\!M$) of pulsar signals together with accurate pulsar distances to map the distribution of ionized gas in the Milky Way.} The dispersion measure is defined as the line of sight integral over the free electron density between the observer and the pulsar,
\begin{equation}
 D\!M = \int\limits_{\mathrm{pulsar}}^{\mathrm{observer}}\!\!\mathrm{d}r\, n_\mathrm{e},
\end{equation}
where $n_\mathrm{e}$ is the three-dimensional free electron density. $D\!M$ can be estimated by measuring the arrival time of a pulse at different frequencies, since the time delay is proportional to $D\!M/\nu^2$. While there is a vast number of known dispersion measures very few of them are complemented by an independent distance estimate. 
The NE2001 model by \cite{cordes-2002} is currently the most popular model for the free electron density of the Milky Way. \corr{It uses 1143 $D\!M$ measurements of which 112 were complemented by distance estimates of varying quality. Additionally it uses 269 pulsar scattering measurements, which only provide very indirect distance constraints.}

In this paper, we \corr{perform nonparametric tomography of a simulation of }the Galactic free electron density from pulsar dispersion measures complemented by independent distance estimates. \corr{By nonparametric tomography we mean a reconstruction with a virtually infinite\footnote{
In numercial practice, the amount of degrees of freedom is the number of pixels used. However, the reconstruction will be resolution independent once the resolution is high enough.} 
number of degrees of freedom using a close to minimal set of prior assumptions that only resolves structures which are supported by the data.
Our assumptions are that the electron density is positive and spatially correlated and that the large-scale electron distribution only shows a variation with distance from the Galactic Centre and height above the Galactic Plane. Both the correlation structure and the scaling behavior have to be inferred from the data. As a consequence, our reconstruction is focused on the large ($\mathrm{kpc}$) scales of the Galactic} free electron density. \corr{Small-scale structures such as HII regions and supernova remnants as well as spiral arms are only recovered if they are sufficiently probed and constrained by the data.}

\corr{Our tomography algorithm is derived from first principles in a Bayesian setting. This has the advantage that all assumptions are clearly states as priors. Additionally, it allows us to provide uncertainty maps of our reconstructions, which are important for any subsequent scientific analysis.}


To get a meaningful map with minimal assumptions, one of course needs a data set of \corr{high} quality. Currently, there are \corr{around} 100 pulsars known with reliable (independent) distance estimates. This only allows for a \corr{nonparametric} reconstruction of the largest features in the Milky Way. \corr{New measurements with the Very Long Baseline Array will soon double the number of pulsars with accurate distances (see \cite{Deller-2011}).} However, with the planned Square Kilometer Array radio interferometer \corr{(SKA)} the number of pulsars with parallax distance estimates might increase to around 10000 (see \cite{Smits-2011}).
In this paper we therefore investigate the feasibility of \corr{nonparametric} tomography of the free electron density and demonstrate the performance of our algorithm by applying it to mock data sets similar to what the SKA might deliver. To that end, we create \corr{four} Galaxy models from the NE2001 code by \cite{cordes-2002} with varying degrees of fluctuations \corr{and contrast} as well as observational mock data sets for up to 10000 pulsars with distance estimates of varying quality and apply our algorithm to these data sets.

The remainder of this paper is \corr{structured} as follows: First, we \corr{derive} our tomography algorithm in Sec.~\ref{sec:algorithm}, explaining our notation, our underlying assumptions as well as all probability density functions involved. Second, we explain our Galaxy models and mock observations in detail in Sec.~\ref{sec:simulation}. In Sec.~\ref{sec:reconstructions}, we compare the electron density distributions reconstructed from mock observations with those from the Galaxy models used to produce them. We summarize our discussion in Sec.~\ref{sec:discussion}.

\section{Reconstruction algorithm}
\label{sec:algorithm}

The reconstruction algorithm applied in this work was derived within the framework of \textit{information field theory} introduced by \cite{IFT-2009}. We also follow -- for most parts -- the notation used by them.
To reconstruct the Galactic free electron density from pulsar dispersion measurements we use a very similar filter formalism \corr{to} the one presented by \cite{Junklewitz-2013}, which in turn is based on the critical filter formalism developed by \cite{Gibbs-2010}, \cite{Crit-2011}, and refined by \cite{niels-smoothness}.

\subsection{Signal model}

In the inference formalism we aim to reconstruct the free electron density field $\rho$, a three-dimensional scalar field. We assume it is related to the observed dispersion measure data $D\!M$ by a linear measurement equation subject to additive and signal independent measurement noise,
\begin{equation}
 D\!M = R\rho + n,\\
\label{eq:data_model}
\end{equation}
where $n$ is the measurement noise and $R\rho$ is the application of the linear response operator $R$ on the field $\rho$,
\begin{equation}
\left(R\rho\right)_i \equiv \int\!\!\mathrm{d}^3x\ R(i,\vec{x})\, \rho(\vec{x}).
\end{equation}
The response operator $R$ describes line-of-sight integrals through the density. It can be defined as
\begin{equation}
 R(i,\vec{x}) = \int\limits_{0}^{\left|\vec{d}_i\right|}\!\!\mathrm{d}r\ \delta\!\left( \vec{x} - r\boldsymbol{\hat{d}}_i \right),
\end{equation}
where $\vec{d}_i$ is the position of pulsar $i$ in a coordinate system centered on Sun and $\delta(\cdot)$ is the three-dimensional Dirac delta-distribution and $\boldsymbol{\hat{d}}_i := \vec{d}_i/|\vec{d}_i|$.

Formally, the free electron density is a continuous field. In practice, we reconstruct a discretized version of this field, e.g.\ a three-dimensional map with some pixel size. One can think of the discretized density field as a vector of dimension $N_{\mathrm{pix}}$ with each component containing the field value in a specific pixel. The \corr{dispersion data} $D\!M$ and the noise $n$ can be regarded as vectors of dimension $N_{\mathrm{data}}$, where each component of $D\!M$ contains a specific measurement result and the corresponding component of $n$ the noise contribution to it. Thus, the response operator becomes a matrix with $N_{\mathrm{pix}}$ columns and $N_{\mathrm{data}}$ rows.

\corr{We parametrize the density as
\begin{equation}
 \rho(\vec{x})  = \Delta(\vec{x}) \tilde{\rho}(\vec{x}),
\end{equation}
where $\Delta$ is the Galactic profile field which describes the the disk shape of the Milky Way.
All deviations from the Galactic profile are described by $\tilde{\rho}$ for which we assume no distinguished direction or position \textit{a~priori}.
To ensure positivity of the density these fields are in turn parametrized as
\begin{equation}
\begin{split}
 \Delta(x,y,z) & = \exp\!\left( \alpha\!\left(\sqrt{x^2+y^2}\right) + \beta\left(|z|\right) \right),\\
 \tilde{\rho}(x,y,z) & = \exp(s(x,y,z)).
 \label{eq:parametrization}
\end{split}
\end{equation}
Thus, $\Delta$ can only represent the vertical and radial scaling behavior of the density and has the degrees of freedom of two one-dimensional functions. On the other hand, $\tilde{\rho}$ retains all degrees of freedom of a three-dimensional field and can represent arbitrary structures. Both, $\Delta$ and $\tilde{\rho}$ are unknown \textit{a~priori} and will} be inferred from the data.

%

\begin{figure}
    \begin{overpic}[width=0.5\textwidth]{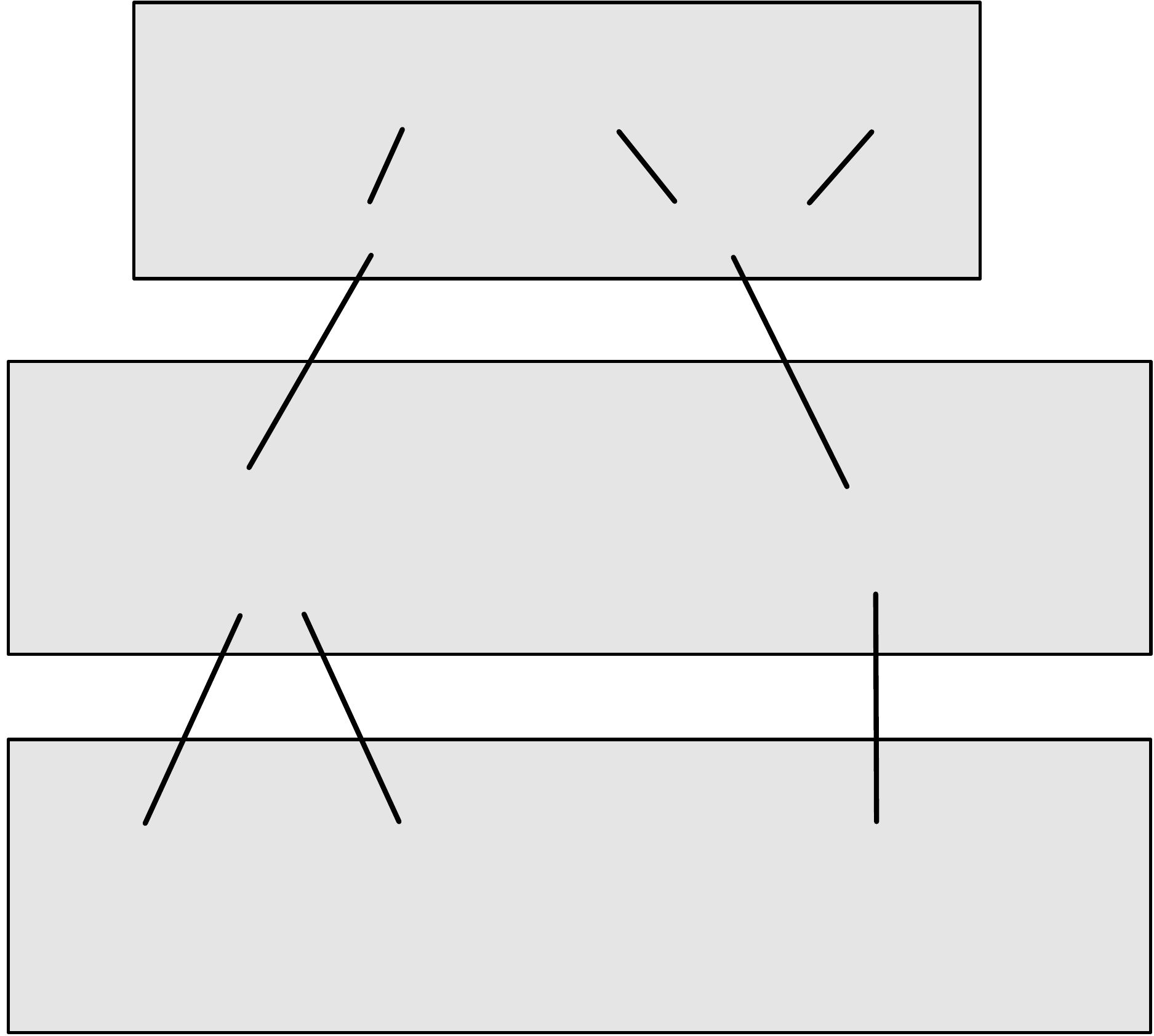}
         \put(36,86){\small{\textbf{parametrization}}}
         \put(13,80){\small{$\ln \rho(x,y,z) = s(x,y,z) + \alpha(\sqrt{x^2\!+\!y^2}) + \beta(|z|)$}}
         \put(25.5,68.5){\small{3D field}}
         \put(57,68.5){\small{1D fields}}
         \put(39,55){\small{\textbf{assumptions}}}
         \put(2,42){\parbox{0.5\linewidth}{\centering
         $s$ is Gaussian with position\\
         and orientation independent\\
         correlation structure}}
         \put(51,42){\parbox{0.5\linewidth}{\centering
         $\alpha$ and $\beta$ are Gaussian\\
         in their second derivatives}}
         \put(36,22){\small{\textbf{implied preferences}}}
         \put(0,10){\parbox{0.25\linewidth}{\centering
         fluctuations\\
         scale with\\
         magnitude}}
         \put(25,10){\parbox{0.35\linewidth}{\centering
         smooth structures\\
         without\\
         sharp edges}}
         \put(62,10){\parbox{0.35\linewidth}{\centering
         exponential\\
         scaling behavior\\
         in $\sqrt{x^2\!+\!y^2}$ and $|z|$}}
    \end{overpic}
    \caption{A diagram outlining the structure of our modeling. 
    }
    \label{fig:model_diagram}
\end{figure}
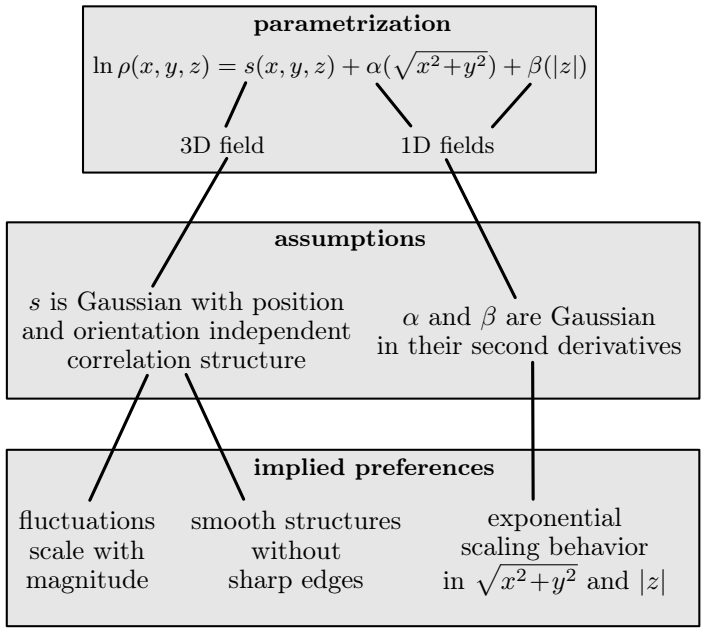

\corr{We summarize our modeling in Fig.~\ref{fig:model_diagram}. The logarithmic density $\rho$ is parametrized by three additive components, one 3D field and two 1D fields. As we outline in Secs.~\ref{sec:prior}~and~\ref{sec:profile} 
all three fields are assumed to follow Gaussian statistics \textit{a priori}. For the 1D fields a specific correlation structure is assumed while the correlation structure of the 3D field is unknown, but assumed to be homogeneous and isotropic. Therefore, our modeling prefers smooth structures, fluctuations that scale with the density, and exponential scaling in radial and vertical directions.
Of course, this is a strong simplification of the Galaxy, where the behaviour of the fluctuations can depend on, e.g., the phase of the interstellar medium or the position within the Galaxy.
However, all of these properties can be recovered if the data demand it, since all degrees of freedom are retained. They are just not part of the prior knowledge entering our inference.}

\subsection{Necessary probability density functions}

Our goal is to derive an algorithm that yields an estimate of the logarithm of the Galactic free electron density. Hence, we construct the posterior probability density function (PDF) $\mathcal{P}(s|\mathrm{data})$, which is the PDF for the signal given the data set $\{D\!M,\vec{d}_\mathrm{obs(erved)}\}$, using Bayes' theorem,
\begin{equation}
 \mathcal{P}(s|\mathrm{data}) = \frac{\mathcal{P}(s,D\!M|\vec{d}_\mathrm{obs})}{\mathcal{P}(D\!M|\vec{d}_\mathrm{obs})} = \frac{\mathcal{P}(s|\vec{d}_\mathrm{obs}) \mathcal{P}(D\!M|s,\vec{d}_\mathrm{obs})}{\mathcal{P}(D\!M|\vec{d}_\mathrm{obs})}.
\label{eq:Bayes}
\end{equation}
On the \corr{right-hand side}, we have three PDFs: the prior $\mathcal{P}(s|\vec{d}_\mathrm{obs}) = \mathcal{P}(s)$, the likelihood $\mathcal{P}(D\!M|s,\vec{d}_\mathrm{obs})$, and the evidence $\mathcal{P}(D\!M|\vec{d}_\mathrm{obs})$.
The evidence is independent from the signal and therefore automatically determined by the normalization of the posterior.
The prior and the likelihood will be addressed in the following sections. \corr{For notational convenience we will drop the dependence on the observed pulsar positions $\vec{d}_\mathrm{obs}$ throughout the rest of this paper.}

\corr{Throughout this section we will assume the Galactic profile field to be given. We will adress its inference in Sec.~\ref{sec:profile}.}

\subsubsection{The likelihood}
\label{sec:likelihood}

The likelihood $\mathcal{P}(D\!M|s)$ is the PDF that an observation yields \corr{dispersion measures} $D\!M$ assuming a specific realization of the underlying signal field $s$. If both the noise $n$ and the pulsar distances $d_i \equiv |\vec{d}_i|$, were known, the relation between the dispersion measure data and signal would be deterministic,
\begin{equation}
 \mathcal{P}(D\!M|s,n,d) = \delta(D\!M-R\rho-n),
\end{equation}
with $\rho(\vec{x}) = \Delta(\vec{x}) \mathrm{e}^{s(\vec{x})}$.
We do not know the realization of the noise, nor do we aim to reconstruct it. It is assumed to follow Gaussian statistics with zero mean and known covariance structure\footnote{We denote expectation values with respect to the underlying PDF as $\left\langle f(x) \right\rangle_{\mathcal{P}(x)} := \int\!\mathcal{D}x\, f(x)\, \mathcal{P}(x)$.},
\begin{equation}
 \left\langle n_i n_j \right\rangle_{\mathcal{P}(n)} = N_{ij} = \delta_{ij} \sigma_i^2,
\end{equation}
where $\sigma_i$ is the \corr{root mean square error} of the observation $i$ and we assumed independent measurements.
Distance information is usually given in the form of parallaxes from which distance estimates can be derived. As all observables these are subject to uncertainties which is why the information about the distances of the pulsars is described by a PDF\footnote{
We assume here that the distance PDF is correctly derived from the parallax PDF taking Lutz-Kelker bias into account (see \cite{Verbiest-2010}).
}, $\mathcal{P}(d) \equiv \mathcal{P}(d|\mathrm{parallaxes})$, which can be non-Gaussian. Since we are doing inference on $s$, we need the noise and distance\footnote{Technically, we also need to marginalize over the position on the sky (i.e. the direction of the line of sight). But since the angular error of the pulsar position is small compared to the error in distance, we can neglect it and treat the direction as an exact value.} marginalized likelihood
\begin{equation}
 \mathcal{P}(D\!M|s) = \int\!\!\mathcal{D}n\mathcal{D}d\ \mathcal{P}(D\!M|s,n,d) \mathcal{P}(n) \mathcal{P}(d),
\label{eq:exact-likelihood}
\end{equation}
where we assumed $n$ and $d$ to be independent from $s$ and each other. \corr{The symbols $\mathcal{D}n$ and $\mathcal{D}d$ denote integration over the full phase space of $n$ and $d$, i.e.~the space of all possible configurations ($\mathcal{D}n \equiv \Pi_i \mathrm{d}n_i$).}

Integration over $n$ in Eq.\ \eqref{eq:exact-likelihood} is trivial and yields
\begin{equation}
 \mathcal{P}(D\!M|s) = \int\!\!\mathcal{D}d\ \mathcal{G}(D\!M-R\rho,N) \mathcal{P}(d),
\end{equation}
\corr{where $\mathcal{G}$ indicates a Gaussian PDF, $\mathcal{G}(x,X) := |2\pi X|^{-\frac{1}{2}} \mathrm{e}^{-\frac{1}{2} x^\dagger X^{-1} x}$. }
Integration over $d$, however, cannot be done analytically, but one can approximate the marginalized likelihood by a Gaussian characterized by its first two moments in $D\!M$. The first moment is
\begin{equation}
 \left\langle D\!M \right\rangle_{\mathcal{P}(D\!M|s)} = \tilde{R}\rho,
\end{equation}
with
\begin{equation}
 \tilde{R}_i(\vec{x}) = \left\langle R_i(\vec{x}) \right\rangle_{\mathcal{P}(d)} = \int\limits_{0}^{\infty}\!\!\mathrm{d}r\ \delta\!\left( \vec{x} - r\boldsymbol{\hat{d}}_i \right)\, P[d_i>r],
\end{equation}
where $P[d_i>r]$ is the probability that the pulsar distance $d_i$ is larger than $r$.
The second moment is
\begin{equation}
 \left\langle D\!M\,D\!M^\dagger \right\rangle_{\mathcal{P}(D\!M|s)} = N+\left\langle \left(R\rho\right)\left(R\rho\right)^\dagger \right\rangle_{\mathcal{P}(d)}.
\end{equation}
For non-diagonal elements the second term on the right hand side decouples,
\begin{equation}
\begin{split}
 \left\langle \left(R\rho\right)_i\left(R\rho\right)_j \right\rangle_{\mathcal{P}(d)} & =  \left\langle \left(R\rho\right)_{i\!\!\phantom{j}} \right\rangle_{\mathcal{P}(d)} \left\langle\left(R\rho\right)_j \right\rangle_{\mathcal{P}(d)}\\
 & = \left( \tilde{R}\rho \right)_i\left( \tilde{R}\rho \right)_j \quad \mathrm{for}\quad i\neq j.
\end{split}
\end{equation}
Diagonal elements yield
\begin{equation}
\begin{split}
 \left\langle \left(R\rho\right)_i\left(R\rho\right)_i \right\rangle_{\mathcal{P}(d)}  = & \int\limits_{\mathbb{R}^3}\!\!\mathrm{d}^3x\int\limits_{\mathbb{R}^3}\!\!\mathrm{d}^3y \ \rho(\vec{x})\rho(\vec{y})\times \\ & \left\langle R_i(\vec{x}) R_i(\vec{y}) \right\rangle_{\mathcal{P}(d)},
\end{split}
\end{equation}
with
\begin{equation}
\begin{split}
 \left\langle R_i(\vec{x}) R_i(\vec{y}) \right\rangle_{\mathcal{P}(d)} = & \int\limits_{0}^{\infty}\mathrm{d}r\int\limits_{0}^{\infty}\mathrm{d}r'\ \delta(\vec{x}-r\boldsymbol{\hat{d}}_i) \delta(\vec{y}-r'\boldsymbol{\hat{d}}_i) \times \\ & P[d_i>\max(r,r')].
\end{split}
\end{equation}

Using these first two moments, we can approximate\footnote{This corresponds to characterizing the likelihood by its cumulants and setting all but the first two cumulants to zero.} the likelihood $\mathcal{P}(D\!M|s)$ by a Gaussian $\mathcal{G}(D\!M-\tilde{R}\rho,\tilde{N})$ with
\begin{equation}
 \tilde{N}_{ii} = N_{ii} + \rho^{\dagger} F^{(i)} \rho,
\label{eq:effective_noise}
\end{equation}
where\footnote{In this work we abbreviate \mbox{$\xi^\dagger \zeta := \int\!\mathrm{d}^3x\, \xi^*(\vec{x})\,\zeta(\vec{x})$} and \mbox{$\Xi\, \xi := \int\!\mathrm{d}^3y\, \Xi(\vec{x},\vec{y})\,\xi(\vec{y})$} for continuous quantities.}
\begin{equation}
\begin{split}
 F^{(i)}(\vec{x},\vec{y}) & := \left\langle R_i(\vec{x})R_i(\vec{y})\right\rangle _{\mathcal{P}(d_i)} -  \tilde{R}_i(\vec{x})\tilde{R}_i(\vec{y}) \\
 &\ = \int\limits_{0}^{\infty}\mathrm{d}r\int\limits_{0}^{\infty}\mathrm{d}r'\ \delta(\vec{x}-r\boldsymbol{\hat{d}}_i) \delta(\vec{y}-r'\boldsymbol{\hat{d}}_i) \times \\  & \quad \ \ P[d_i>\max(r,r')]P[d_i<\min(r,r')].
\end{split}
\end{equation}

The noise covariance matrix of this effective likelihood is signal dependent, which increases the complexity of the reconstruction problem.
Therefore, we approximate the density in Eq.\ \eqref{eq:effective_noise} by its posterior mean,
\begin{equation}
 \rho^{\dagger} F^{(i)} \rho = \mathrm{tr}\left( \rho \rho^\dagger F^{(i)} \right) \approx \mathrm{tr}\left( \left\langle\rho\right\rangle_{\mathcal{P}(\rho|D\!M)} \left\langle\rho\right\rangle_{\mathcal{P}(\rho|D\!M)}^\dagger F^{(i)} \right).
\label{eq:noise_addition}
\end{equation}
Since $\left\langle\rho \right\rangle_{\mathcal{P}(\rho|D\!M)}$ depends on $\tilde{N}$ this yields a set of equations that need to be solved self-consistently (see Sec.~\ref{sec:filter_equations}).

\subsubsection{The priors}
\label{sec:prior}

The signal field $s$ is unknown \textit{a~priori}, but we assume that it has some correlation structure. We describe this correlation structure by moments up to second order in $s$. The principle of maximum entropy therefore \corr{requires} that our prior probability distribution has a Gaussian form,
\begin{equation}
 \mathcal{P}(s|S) = \mathcal{G}(s,S) := \left| 2\pi S \right|^{-\frac{1}{2}} \exp\!\left( -\frac{1}{2} s^\dagger S^{-1} s \right),
\end{equation}
with some unknown correlation structure,
\begin{equation}
 S(\vec{x},\vec{y}) = \left\langle s(\vec{x}) s(\vec{y}) \right\rangle_{\mathcal{P}(s)}.
\end{equation}
The first moment of $s$ is set to zero, since it can be absorbed into $\Delta(\vec{x})$. So the \textit{a~priori} mean of $s$ is contained in $\Delta(\vec{x})$.

\textit{A~priori}, our algorithm has no \corr{preferred} direction or position for $s$. This reduces the number of degrees of freedom of the correlation structure $S$. It is fully described by a power spectrum $p(k)$,
\begin{equation}
 S(\vec{x},\vec{y}) = \sum\limits_k\,S^{(k)}(\vec{x},\vec{y})\, p(k),
\end{equation}
where $S^{(k)}$ is the projection operator onto the spectral band $k$ with its Fourier transform defined as
\begin{equation}
 S^{(k)}(\vec{q},\vec{q'}) = (2 \pi)^3 \delta(\vec{q} - \vec{q'}) \mathbb{1}_k\!\left(|\vec{q}|\right),
\end{equation}
with
\begin{equation}
 \mathbb{1}_k\!\left(|\vec{q}|\right) = \begin{cases}
                                         1 & \mathrm{for}\ \  |\vec{q}|=k \\
					0 & \mathrm{otherwise}
                                        \end{cases}.
\end{equation}

The power spectrum $p(k)$, however, is still unknown. The prior for the power spectrum is constructed out of two parts. First, an inverse Gamma distribution $\mathcal{I}(p(k);\alpha_k,q_k)$ for each $k$-bin (see Appendix~\ref{app:parameters}), which is a conjugate prior for a Gaussian PDF, second a Gaussian cost-function that punishes deviations from power law spectra (see \cite{niels-smoothness}),
\begin{equation}
 \mathcal{P}(p) \propto \left\{\prod_k \mathcal{I}(p(k);\alpha_k,q_k)\right\}  \exp\!\left( -\frac{1}{2} (\log p)^\dagger T (\log p) \right).
\end{equation}
$T$ is an operator that fulfills
\begin{equation}
 (\log p)^\dagger T (\log p) = \frac{1}{\sigma_p^2} \int\!\!\mathrm{d}(\log k) \left(\frac{\partial^2\log p(k)}{\partial (\log k)^2} \right)^2,
 \label{eq:smoothness-prior}
\end{equation}
and $\sigma_p$ is a parameter that dictates how smooth the power spectrum is expected to be. In \corr{our paper} $\log$ refers to the natural logarithm.
We explain our choice of the parameters $\alpha_k$, $q_k$, and $\sigma_p$ in Appendix~\ref{app:parameters}.

\subsubsection{The power spectrum posterior}
\label{sec:p_prior}

With the signal and power spectrum priors, we can derive a posterior for the power spectrum,
\begin{equation}
\begin{split}
 \mathcal{P}(p|D\!M) & \propto \int\!\!\mathcal{D}s\ \mathcal{P}(D\!M|s,p)\,\mathcal{P}(s|p)\,\mathcal{P}(p)\\
 & = \int\!\!\mathcal{D}s\ \mathcal{P}(D\!M|s)\,\mathcal{G}(s,S)\,\mathcal{P}(p).
\end{split}
\end{equation}
We calculate the integral using a saddle point approximation up to second order around the maximum for the $s$-dependent part,
\begin{equation}
\begin{split}
 \mathcal{P}(D\!M|s)\,\mathcal{G}(s,S) & \approx \mathcal{P}(D\!M|m)\,\mathcal{G}(m,S)\,\mathrm{e}^{-\frac{1}{2}(s-m)^\dagger D^{-1}(s-m)}\\
 & \propto \mathcal{G}(m,S)\,\mathrm{e}^{-\frac{1}{2}(s-m)^\dagger D^{-1}(s-m)}
\end{split}
\end{equation}
where $m$ and $D$ are defined as $m^{(s)}$ and $D^{(s)}$ in Sec.\ \ref{sec:posterior} and only $s$ and $p$-dependent factors are kept after the proportionality sign.
With this approximation we arrive at
\begin{equation}
 \mathcal{P}(p|D\!M) \propto \left| 2\pi D \right|^{\frac{1}{2}}\,\left| 2\pi S \right|^{-\frac{1}{2}}\,\mathrm{e}^{-\frac{1}{2}m^\dagger S^{-1} m}\, \mathcal{P}(p).
\end{equation}
Maximizing this PDF with respect to $\log(p)$ (see \cite{niels-smoothness}) leads to
\begin{equation}
 p(k) = \frac{q_k + \frac{1}{2}\mathrm{tr}\!\left(S^{(k)}(mm^\dagger + D)   \right)}{\alpha_k - 1 +\frac{1}{2}\varrho_k+(T\log p)_k} ,
\label{eq:powspec_approx}
\end{equation}
where $\varrho_k = \mathrm{tr}\left(S^{(k)}\right)$ is the number of degrees of freedom in the spectral band $k$.
This formula for the power spectrum $p(k)$ should be solved self-consistently, since $m$ and $D$ depend on $p(k)$ as well. Thus we arrive at an iterative scheme, where we look for a fixed point of Eq.\ \eqref{eq:powspec_approx}.

\subsubsection{The signal posterior}
\label{sec:posterior}

The signal posterior can be expressed as
\begin{equation}
 \mathcal{P}(s|D\!M) = \int\!\!\mathcal{D}(\log p)\, \mathcal{P}(\log p|D\!M)\, \mathcal{P}(s|p,D\!M),
\end{equation}
where $\mathcal{P}(s|p,D\!M)$ is the signal posterior with a given power spectrum. Instead of calculating the marginalization over $\log p$ we use Eq.~\eqref{eq:powspec_approx} for the power spectrum; i.e., we approximate $\mathcal{P}(\log p|D\!M)$ by a Dirac peak at its maximum. This procedure is known as the Empirical Bayes method.
The signal posterior with a given power spectrum is proportional to the product of the signal prior and the likelihood (see Eq.\ \eqref{eq:Bayes}),
\begin{equation}
\begin{split}
 \mathcal{P}(s,D\!M|p,\tilde{N}) \propto & \exp\!\left( -\frac{1}{2} s^\dagger S^{-1} s \right)\times\\ & \exp\!\left( -\frac{1}{2} (D\!M-\tilde{R}\rho)^\dagger \tilde{N}^{-1} (D\!M-\tilde{R}\rho) \right)
\end{split}
\end{equation}
However, as has been demonstrated in Sec.\ \ref{sec:p_prior} and \ref{sec:likelihood}, $S$ and $\tilde{N}$ depend on the mean and the covariance of $\mathcal{P}(s|D\!M,p,\tilde{N})$ leading to a circular dependence that needs to be solved self-consistently.

We approximate the mean of the posterior by minimizing the joint Hamiltonian $\mathcal{H}(s,D\!M|p,\tilde{N}) := -\log \mathcal{P}(s,D\!M|p,\tilde{N})$ with respect to $s$,
\begin{equation}
 m^{(s)} \approx \underset{s}{\mathrm{arg\,min}}\ \mathcal{H}(s,D\!M|p,\tilde{N}),
\end{equation}
and its covariance by the inverse Hessian at that minimum,
\begin{equation}
 D^{(s)} \approx \left(\left. \frac{\delta^2}{\delta s \delta s^\dagger} \mathcal{H}(s,D\!M|p,\tilde{N}) \right|_{s=m}\right)^{-1}.
 \label{eq:inverse_Hessian}
\end{equation}
These estimates are the \textit{maximum~a~posteriori} (MAP) estimates of $s$.
Consequently $m^{(\rho)}$ and $D^{(\rho)}$ are estimated as
\begin{equation}
 m^{(\rho)}(\vec{x}) \approx \Delta(\vec{x})\mathrm{e}^{m^{(s)}(\vec{x})}
\end{equation}
and
\begin{equation}
\begin{split}
 D^{(\rho)}(\vec{x},\vec{y}) & \approx \Delta(\vec{x})\mathrm{e}^{m^{(s)}(\vec{x})}\left( \mathrm{e}^{D^{(s)}(\vec{x},\vec{y})} - 1 \right)\mathrm{e}^{m^{(s)}(\vec{y})}\Delta(\vec{y}).\\
\end{split}
\end{equation}
$S$ is then constructed as
\begin{equation}
 S(\vec{x},\vec{y}) = \sum\limits_k\,S^{(k)}(\vec{x},\vec{y})\, p(k),
\end{equation}
with $p(k)$ given by Eq.~\eqref{eq:powspec_approx}.
$\tilde{N}$ is constructed using Eqs.~\eqref{eq:effective_noise} and \eqref{eq:noise_addition} as
\begin{equation}
 (\tilde{N})_{ij} = (N)_{ij} + \delta_{ij}\, \mathrm{tr}\left(  m^{(\rho)} m^{(\rho)\dagger}  F^{(i)} \right),
\end{equation}
where $\delta_{ij}$ is the Kronecker delta.

\subsection{Galactic profile inference}
\label{sec:profile}

To infer the Galactic profile field $\Delta$ we introduce $\tilde{s} \equiv s + \log(\Delta) \equiv \log(\rho)$. The Galactic profile is to capture the most prominent symmetries of a disk galaxy, namely its rotational symmetry and the scaling \corr{behaviour with radial distance from the Galactic center and} vertical distance from the \corr{Galactic plane.
Using Eq.~\eqref{eq:parametrization} $\mu \equiv \log(\Delta)$ becomes}
 \begin{equation}
  \mu(x,y,z) = \alpha(r) + \beta(|z|),\quad \mathrm{with} \quad r\equiv\sqrt{x^2+y^2},
   \label{eq:profiles}
 \end{equation}
where $\alpha$ and $\beta$ are one-dimensional functions describing the average behavior with respect to the radial distance and the vertical distance from the Galactic center.
\corr{Including the shift by $\mu$ from $s$ to $\tilde{s}$} yields the signal prior
\begin{equation}
 \mathcal{P}(\tilde{s}|\alpha,\beta) = \mathcal{G}(\tilde{s}-\mu,S).
\end{equation}
We do not want to assume specific functions $\alpha$ and $\beta$ but to infer them. To that end we choose a Gaussian prior,
\begin{equation}
 \mathcal{P}(\alpha,\beta) \propto \exp\!\left(- \frac{1}{2 \sigma_\alpha^2} \left(\frac{\partial^2 \alpha}{\partial r^2} \right)^2 - \frac{1}{2 \sigma_\beta^2} \left(\frac{\partial^2 \beta}{\partial |z|^2} \right)^2 \right),
\end{equation}
with the second derivative of $\alpha$ (or $\beta$ respectively) as the argument. This prior prefers linear functions for $\alpha$ and $\beta$ and thus Galactic profile fields with an exponential fall-off (or rise).
To simplify the notation we define $\xi(r,|z|) = \left(\alpha(r),\beta(|z|)\right)^T$ and introduce the linear operators $\Xi$ and $X$, where
\begin{equation}
 X\xi = \alpha + \beta \equiv \mu
\label{eq:X_operator}
\end{equation}
and
\begin{equation}
 \xi^\dagger \Xi \xi = \frac{1}{\sigma_\alpha^2} \left(\frac{\partial^2 \alpha}{\partial r^2} \right)^2 + \frac{1}{\sigma_\beta^2} \left(\frac{\partial^2 \beta}{\partial |z|^2} \right)^2.
\end{equation}

Now we can write the Hamiltonian of $\xi$ given a specific electron density as
\begin{equation}
\begin{split}
 \mathcal{H}(\xi|\tilde{s}) & = \frac{1}{2} \left( \tilde{s} - X\xi\right)^\dagger S^{-1} \left( \tilde{s} - X\xi\right) + \frac{1}{2} \xi^\dagger \Xi \xi + \mathrm{const.} \\
 & = \frac{1}{2} \xi^\dagger \left( X^\dagger S^{-1} X + \Xi \right) \xi - \tilde{s}^\dagger S^{-1} X \xi + \mathrm{const.} \\
 & \equiv \frac{1}{2} \xi^\dagger D_{(\xi)}^{-1} \xi - j^\dagger_{(\xi)} \xi + \mathrm{const.}
 \label{eq:profile_Hamiltonian}
\end{split}
\end{equation}
with $D_{(\xi)} = \left( X^\dagger S^{-1} X + \Xi \right)^{-1}$ and $j_{(\xi)} =  X^\dagger S^{-1} \tilde{s}$. 
Since this Hamiltonian is a quadratic form in $\xi$ the mean of the corresponding Gaussian PDF is
\begin{equation}
 \left\langle \xi \right\rangle_{(\xi|\tilde{s})} = D_{(\xi)} j_{(\xi)}
\end{equation}

\subsection{Filter equations}
\label{sec:filter_equations}

Using the posterior estimates presented in the previous section, we arrive at the following iterative scheme to reconstruct the density $\rho$:
\begin{enumerate}
 \item Make an initial guess for the power spectrum (e.g.\ some power law) and the additive term in the noise covariance (e.g.\ simple relative error propagation).
 \item With the current estimates for $p$ and $\tilde{N}$ the Hamiltonian, \mbox{$\mathcal{H}(s,D\!M|p,\tilde{N}) + \mathrm{const.} \equiv \log\mathcal{P}(s,D\!M|p,\tilde{N})$}, is
\begin{equation}
\begin{split}
 \mathcal{H}(s,D\!M|p,\tilde{N}) & = \frac{1}{2} s^\dagger \left( \sum\limits_k S^{(k)} p_k^{-1} \right) s \\
 &\quad + \frac{1}{2}\left(\mathrm{e}^s * \Delta \right)^\dagger \tilde{R}^\dagger \tilde{N}^{-1}\tilde{R} \left(\mathrm{e}^s * \Delta \right)\\
 &\quad - D\!M^\dagger \tilde{N}^{-1} \tilde{R} \left(\mathrm{e}^s * \Delta \right),
\end{split}
\end{equation}
where $*$ denotes point-wise multiplication in position space.
\label{final-filter1}
 \item The MAP estimate of this Hamiltonian is calculated as
\begin{equation}
  m^{(s)} = \underset{s}{\mathrm{arg\,min}}\ \mathcal{H}(s,D\!M|p,\tilde{N}),
\end{equation}
 with the covariance estimate (see Appendix~\ref{app:derivatives})
\begin{equation}
 D^{(s)} = \left( \left. \frac{\delta^2}{\delta s \delta s^\dagger} \mathcal{H}(s,D\!M|p,\tilde{N}) \right|_{s=m} \right)^{-1}.
\label{eq:cov_estimate}
\end{equation}
\label{final-filter2}
 \item The updated power spectrum is the solution (with respect to $p(k)$) of the equation
\begin{equation}
 p(k) = \frac{q_k + \frac{1}{2}\mathrm{tr}\!\left(S^{(k)}(mm^\dagger + D)   \right)}{\alpha_k - 1 +\frac{1}{2}\varrho_k+(T\log p)_k}.
\end{equation}
\label{final-filter3}
 \item The updated effective noise covariance is calculated as
\begin{equation}
 (\tilde{N})_{ii} = (N)_{ii} + \mathrm{tr}\left(  m^{(\rho)} m^{(\rho)\dagger}  F^{(i)} \right),
\end{equation}
with
\begin{equation}
\begin{split}
 m^{(\rho)}(\vec{x}) & = \Delta(\vec{x})\exp\!\left(m^{(s)}(\vec{x})\right).\\
\end{split}
\end{equation}
\label{final-filter4}

 \item The updated Galactic profile field is
 \begin{equation}
 \begin{split}
  \Delta & = \exp\!\left(X m^{(\xi)}\right)\qquad \mathrm{with} \\
  m^{(\xi)} & = \left( X^\dagger S^{-1} X + \Xi \right)^{-1} X^\dagger S^{-1} \log(m^{(\rho)})
  \end{split}
 \end{equation}

 \item Repeat from step \ref{final-filter1} until convergence is reached.
\end{enumerate}

\vspace{0.2cm}
When the solution of this set of equations is converged, the estimate of the density $\rho$ is
\begin{equation}
 \rho(\vec{x}) \approx m^{(\rho)}(\vec{x}) \pm \sigma^{(\rho)}(\vec{x}),
\end{equation}
where the confidence interval $\sigma^{(\rho)}$ is defined as
\begin{equation}
 \sigma^{(\rho)}(\vec{x}) := \sqrt{m^{(\rho)}(\vec{x})\left( \mathrm{e}^{D^{(s)}(\vec{x},\vec{x})} - 1 \right)m^{(\rho)}(\vec{x})}.
\label{eq:uncertainty}
\end{equation}

\section{Application to simulated data}
\label{sec:simulation}

To test the reconstruction of the Galactic free electron density distribution with the SKA we generate mock data sets of \corr{pulsars with various distance} uncertainties. We simulate pulsar \corr{populations} using the PSRPOPpy package by \cite{Bates-2014}, which is based on the pulsar population model by \cite{Lorimer-2006}. \corr{The generated populations take into account the observational thresholds of the SKA (mid-frequency). These data sets} sample modified versions of the NE2001 model by \cite{cordes-2002} \corr{through dispersion measures.}


\subsection{Galaxy model}
\label{sec:galaxy_model}

\begin{figure}
 \includegraphics[width=\linewidth]{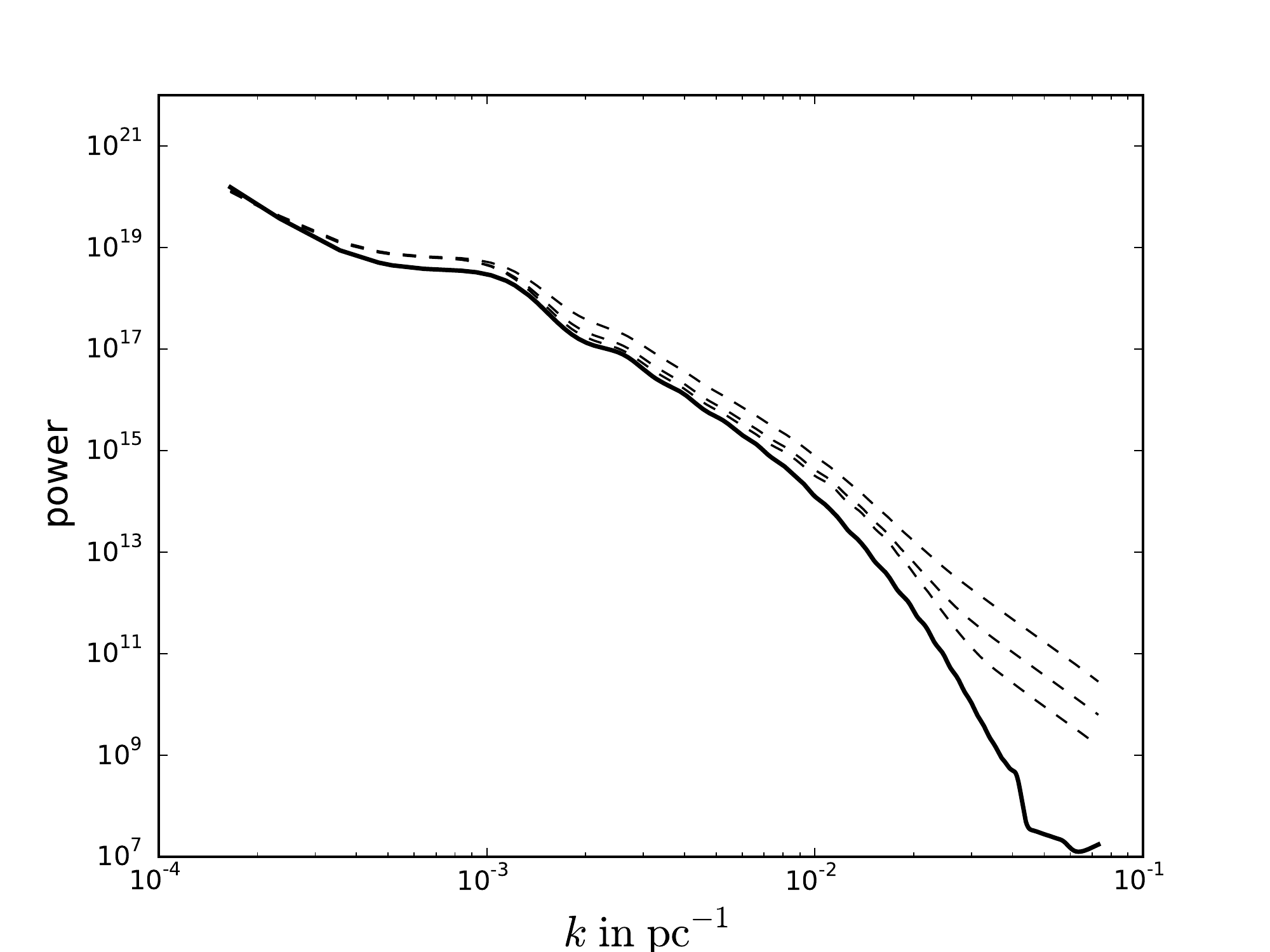}
 \caption{The power spectrum of the NE2001 field without local features, clumps, and voids compared to the three \corr{unenhanced} Galaxy models. The thick solid line depicts the NE2001 spectrum, the thin dashed lines depict the spectra of the models with strong, medium and weak (from top to bottom) fluctuations. For the calculation of the spectra, the density peak in the Galactic center is masked.}
 \label{fig:fluct_power}
\end{figure}

We deactivated\footnote{
This is achieved by modifying the ``nelism.inp'', ``neclumpN.NE2001.dat'', and ``nevoidN.NE2001.dat'' files provided with the \corr{NE2001} code.
}
all local ISM components as well as all clumps and voids in the NE2001 model. \corr{We keep the clump in the Galactic center, since it is the only one at a distinguished position.}
We evaluated\footnote{
To get the three-dimensional free electron density from the compiled NE2001 code, we evaluate two positions in each pixel, that have parallel line-of-sight vectors. The difference between their dispersion measures divided by the difference of their distance to Sun is then taken as the free electron density in that pixel.
} 
the \corr{resulting free electron density model} in a 512x512x64 pixel grid centered on the Galactic center with a pixel edge length of $75\,\mathrm{pc}$. This means that our model \corr{extends out to} $2400\,\mathrm{pc}$ from the Galactic plane. We assume a density of zero outside of this regime when calculating the dispersion measures.
The resulting density field is very smooth. 
\corr{We generate three Gaussian random fields which follow a power-law distribution with a spectral index\footnote{
There is no physical reason for that choice, but a power law with this index seems to follow the spectrum of the log-density in the original model NE2001 rather well on large to medium scales.
} of -4.66 but have different fluctuation amplitudes. We make sure that the} Sun sits in an underdensity \corr{in these random fields. Then we add these three random field maps to our smooth map of $\log(n_\mathrm{e})$ to create three different modified versions of NE2001.}
In Fig.~\ref{fig:fluct_power} we depict the power spectra of the smooth NE2001 field (without local features, clumps and voids) and the power spectra of the \corr{three modified versions of it.}

\subsubsection{Contrast enhanced model}
\label{sec:enhanced_contrast}

\corr{
The three Galaxy models we generated from NE2001 have relatively little contrast in the sense that under- and overdense regions differ by relatively moderate factors. For example, the density in the region between the Perseus and the Carina-Sagittarius arm where the Sun is located is only a factor of three lower than in the Perseus arm itself. Since the Perseus arm is a less than $1\,\mathrm{kpc}$ in width any excess dispersion measure due to the arm can also be explained by an underestimated pulsar distance for many lines of sight. In consequence, we expect the reconstruction quality to improve if the input model has higher contrast. Therefore, we prepare one additional model with enhanced contrast. To that end, we take the input model with medium strength fluctuations as described above. We divide out the scaling behavior in radial and vertical directions using the scale heights from NE2001. We square the density and divide it by a constant to ensure that the mean density in the Galactic plane remains unchanged}\footnote{
\corr{The bulge in the Galactic center is kept unchanged by the whole procedure.}
}\corr{. Finally we multiply the resulting density with the scaling functions to restore the original scaling in radial and vertical directions.}

\corr{This procedure yields a Galaxy model sharing the same morphology and scaling behavior as the input model. Averaged over the lines-of-sight, the value of dispersion measures is roughly unchanged.
But the contrast is twice as strong, i.e., the previously mentioned factor between the density in the Perseus arm and the inter-arm region is now squared from 3 to 9. We will show a picture of the density in the Galactic plane of this model in Sec.~\ref{sec:midplane_comp}, where we compare it with its reconstruction.}

\subsection{Simulated population and survey}
\label{sec:mock_population}

\begin{figure}
 \includegraphics[width=0.46\textwidth]{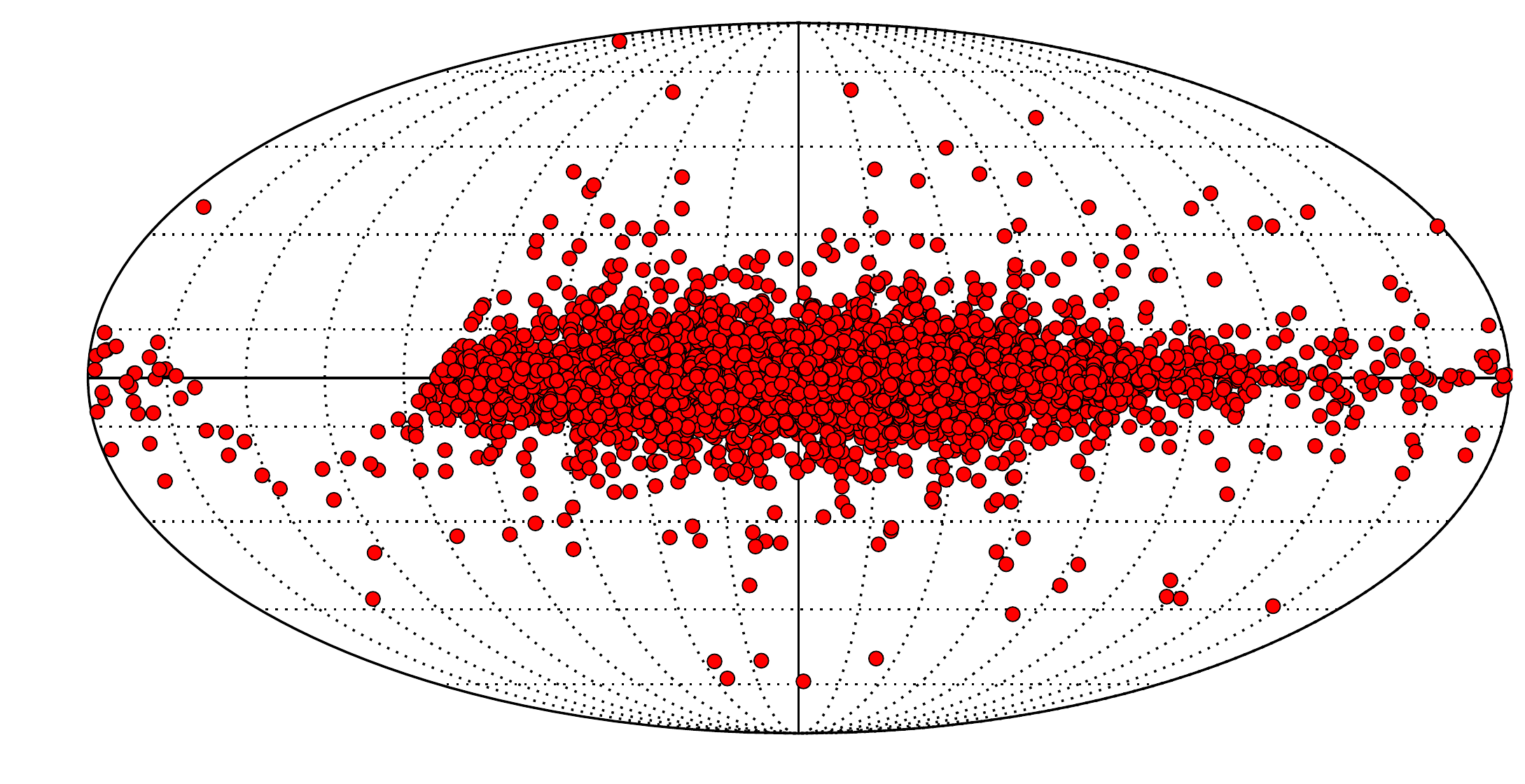}
 \caption{The positions of the simulated 10000 pulsars on the sky in Galactic coordinates.}
 \label{fig:sky_10000}
\end{figure}

We use the ``SKA'' template in the PSRPOPpy package, but reduce the maximum declination in equatorial coordinates to $50^\circ$ (due to the SKAs position on the Southern Hemisphere, see e.g.~\cite{Smits-2009}). This yields a detected population of roughly 14000 pulsars. Out of these, we take the first 1000, 5000, or 10000 pulsars as our test populations. The population is not ordered in any sense, so the first, e.g., 1000 pulsars represent a random sample from the whole detected population. In reality, Malmquist bias will select preferentially pulsars that lie close to the Sun. We choose, however, a random selection in order to see the effect of the population size on the quality of the reconstruction more clearly. In Fig.~\ref{fig:sky_10000} we depict the population of 10000 pulsars projected onto the sky. The pulsars are concentrated towards the center of the Galaxy. The gap in the equatorial Northern Hemisphere is clearly evident in the left part of the plot.

\subsection{Simulated dispersion measures and distances}
\label{sec:mock_data}

We calculate the line integrals through the Galaxy models from the positions generated by the PSRPOPpy package to Sun to generate simulated dispersion measures. 
We add Gaussian random variables to the pulsar distances to simulate measurement uncertainties \corr{of the distances}; for each pulsar we generate one random number and scale this to $5\%$, $15\%$, or $25\%$ of the distance of the pulsar. 
\corr{In reality the distance PDF would be non-Gaussian. The exact form depends on the combination of observables which are used to infer the distance. We use Gaussian PDFs to keep things simple. As long as the real distance PDFs are unimodal we do not expect this choice to have a significant effect on our study.}
We do not simulate additional measurement noise \corr{for the dispersion measures}, as it is expected to be small compared to the distance uncertainty. This leaves us with a number of data sets described in Table~\ref{table:data_sets}. As can be seen in this table we omit the combinations of $1000$ pulsars at $25\%$ distance error (as we do not hope for a good reconstruction in that case) and $10000$ pulsars at $5\%$ distance error (as we deem it to be too unrealistic). 

\begin{table}
\caption{The types of data sets simulated for all Galaxy models. The columns indicate the number of pulsars, the rows the relative distance uncertainties.}              
\label{table:data_sets}      
\centering                                      
\begin{tabular}{r c c c}          
\hline\hline
   & 1000 pulsars & 5000 pulsars & 10000 pulsars \\    
\hline                                   
 $25\%$ unc. &  & \checkmark & \checkmark \\
 $15\%$ unc. & \checkmark & \checkmark & \checkmark \\
  $5\%$ unc. & \checkmark & \checkmark & \\
\hline
\end{tabular}
\end{table}

\begin{figure}
 \includegraphics[width=0.49\textwidth]{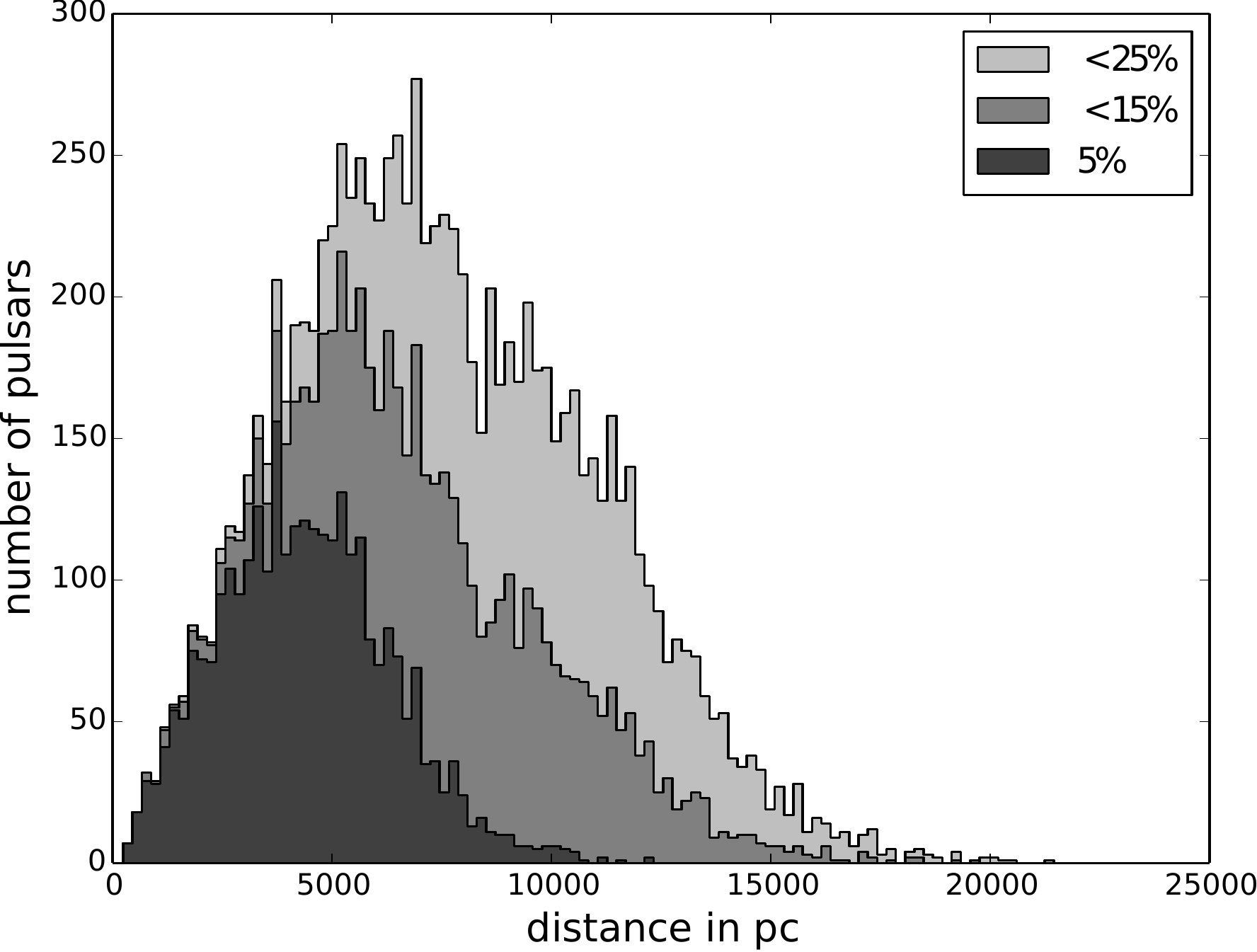}
 \caption{A histogram showing distribution of distance uncertainties with respect to the distance from Sun in the mixed measurement set.}
 \label{fig:comb-hist}
\end{figure}

The aforementioned measurement scenarios are chosen to see the effect of the population size and the distance error on the reconstruction in isolation. A more realistic setting is of course a mix of distance uncertainties where more distant pulsars on average have larger distance errors. We therefore create one additional measurement scenario for 10000 pulsars, where we assign the uncertainty magnitude of each pulsar randomly\footnote{Each pulsar is assigned probabilities to belong to either the $5\%$, the $15\%$ or the $25\%$ set. The probabilities depend on its distance, making more distant pulsars more likely to have higher uncertainties. The pulsar is then randomly assigned to an uncertainty set according to the probabilities.}. The distance uncertainties are distributed as shown in Fig.~\ref{fig:comb-hist}. In this measurement set, 2969 pulsars have a $5\%$ distance error, 3400 pulsars have a $15\%$ distance error, and 3631 pulsars have a $25\%$ distance error. Throughout the rest of this paper we refer to this data set as the ``mixed data set''.

\corr{A very rough estimate of the scales that we can hope to resolve is given by mean distance between neighboring pulsars and the average misplacement due to distance errors. The mean distance between neighboring pulsars is $490\,\mathrm{pc}$ for 1000, $290\,\mathrm{pc}$ for 5000, and $230\,\mathrm{pc}$ for 10000 pulsars. The average misplacement is $380\,\mathrm{pc}$ for $5\%$, $1100\,\mathrm{pc}$ for $15\%$, and $1900\,\mathrm{pc}$ for $25\%$ distance errors and $1300\,\mathrm{pc}$ for the mixed data set. Interpreting these distances as independent uncertainties one can combine them by adding the squares and taking the square root. This provides us with a rough estimate of sampling distances. In Table~\ref{table:distances} we list these distances for each data set.}

\begin{table}
\caption{The estimated sampling distances for each data set. The columns indicate the number of pulsars, the rows the relative distance uncertainties.}              
\label{table:distances}      
\centering                                      
\begin{tabular}{r r r r}          
\hline\hline
   & 1000 pulsars & 5000 pulsars & 10000 pulsars \\    
\hline                                   
 $25\%$ unc. &  & $1900\,\mathrm{pc}$ & $1900\,\mathrm{pc}$ \\
 $15\%$ unc. & $1200\,\mathrm{pc}$ & $1100\,\mathrm{pc}$ & $1100\,\mathrm{pc}$ \\
  $5\%$ unc. & $600\,\mathrm{pc}$ & $500\,\mathrm{pc}$ & \\
\hline
\end{tabular}
\end{table}

\subsection{Algorithm setup}
\label{sec:setup_algorithm}

The algorithm is set up in a $128 \times 128 \times 48$ pixel grid centered on the Galactic center with pixel dimensions\footnote{
\corr{We note that the pixels of our algorithm setup are significantly larger than those of the input models. This is on purpose, since in reality there will always be structure smaller than the chosen pixel size.}
} of $281.25\,\mathrm{pc} \times 281.25\,\mathrm{pc} \times 250\,\mathrm{pc}$.
\corr{While the dispersion measures in our data sets are free from instrumental noise, it is assumed to be $2\%$ in the algorithm. This provides a lower limit for the effective noise covariance (Eq.~\ref{eq:effective_noise}) and thus ensures stability of the inference without losing a significant amount of precision.}
The initial guess for the power spectrum is a broken power law with an exponent of $-3.66$\footnote{\corr{We could in principle use any power spectrum as an initial guess. The choice here comes from no particular reasoning. It has negligible influence on the final result (see Appendix~\ref{sec:convergence}).}}. For the propagated distance uncertainty it is 
\begin{equation}
\sigma_i = \frac{\sqrt{\mathrm{Var}[d_i]}}{d_i} D\!M_i, 
\end{equation}
where $d_i$ is the distance of the pulsar (see Sec.~\ref{sec:likelihood}). \corr{The} initial guesses of the Galactic profile functions\footnote{
We note that while the priors for the profile functions prefer linear forms, all functional forms are allowed in principle.
} are
\begin{equation}
\alpha(r) = \frac{-r}{28000\,\mathrm{pc}}\quad \mathrm{and} \quad \beta(|z|) = \frac{-|z|}{1600\,\mathrm{pc}}.
\end{equation}
We discuss the convergence and final values of the power spectrum, effective errors, and profile functions in Appendix~\ref{sec:convergence}.

\section{Simulation evaluation}
\label{sec:reconstructions}

\corr{Our algorithm accounts for most of the variance in the data while regularizing the result to avoid overfitting. Most of the reconstructions shown in this section have corresponding reduced $\chi^2$ values close to 1, indicating that they show all structures which are sufficiently constained by the data.
We discuss the reduced $\chi^2$ values in detail in Appendix~\ref{sec:chisquared}.}

\subsection{Density in the \corr{midplane}}
\label{sec:midplane_comp}
\begin{figure*}[p]
 \begin{tabular}{c c c}
        \begin{overpic}[width=0.33\textwidth]{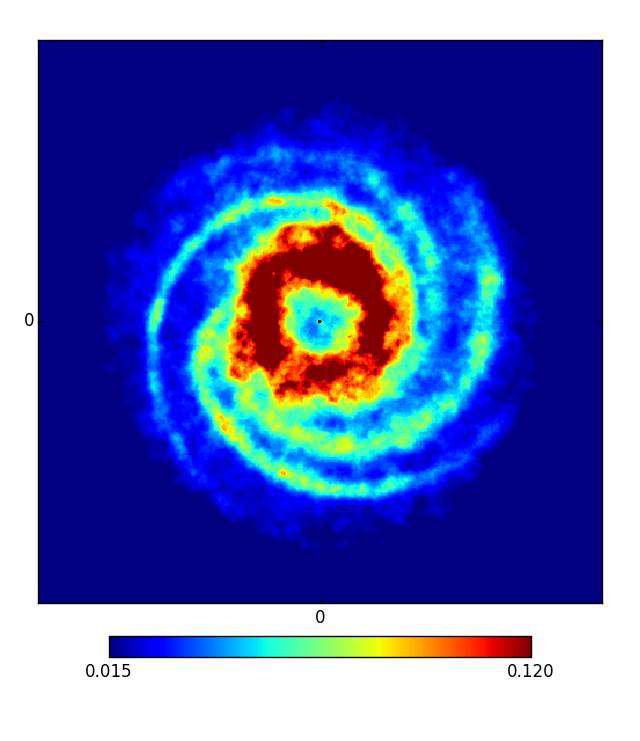}
            \put(35,87){\boldmath\color{white}\textbf{original}}
	    \put(24.1,55.65){\boldmath\color{white}$\bullet$}
	    \put(24.1,55.65){\boldmath\color{black}$\circ$}
        \end{overpic}
     &
     \begin{overpic}[width=0.33\textwidth]{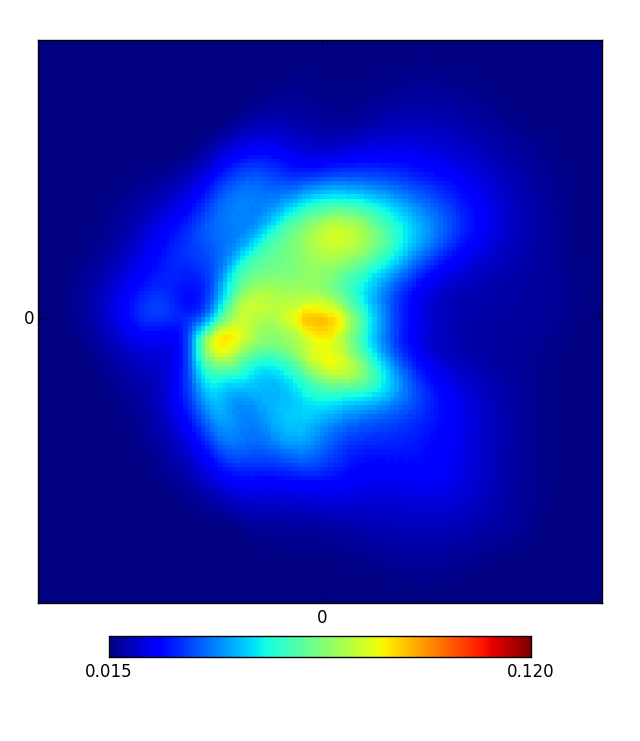}
         \put(33,87){\boldmath\color{white}$5000$ $25\%$}
         \put(24.1,55.65){\boldmath\color{white}$\bullet$}
         \put(24.1,55.65){\boldmath\color{black}$\circ$}
     \end{overpic}
     &
     \begin{overpic}[width=0.33\textwidth]{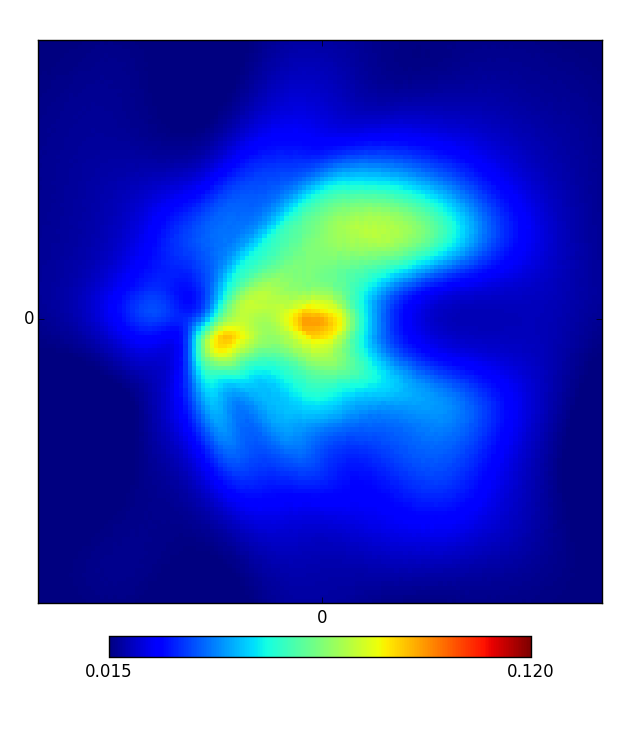}
         \put(30,87){\boldmath\color{white}$10000$ $25\%$}
         \put(24.1,55.65){\boldmath\color{white}$\bullet$}
         \put(24.1,55.65){\boldmath\color{black}$\circ$}
     \end{overpic}
  \\
     \begin{overpic}[width=0.33\textwidth]{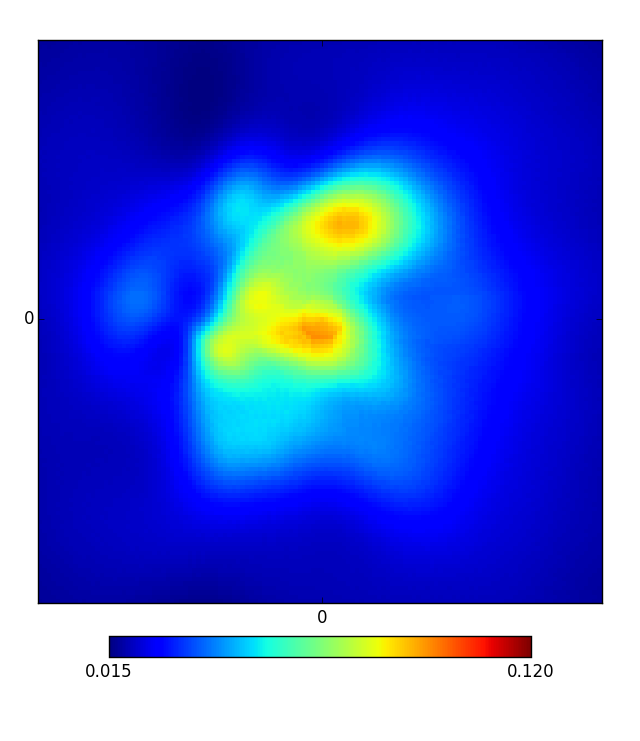}
         \put(33,87){\boldmath\color{white}$1000$ $15\%$}
         \put(24.1,55.65){\boldmath\color{white}$\bullet$}
         \put(24.1,55.65){\boldmath\color{black}$\circ$}
     \end{overpic}
     &
     \begin{overpic}[width=0.33\textwidth]{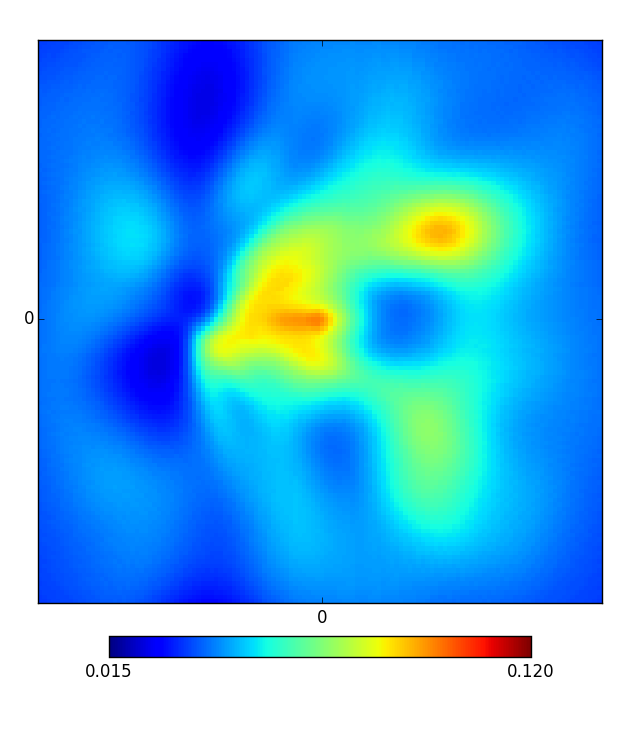}
         \put(33,87){\boldmath\color{white}$5000$ $15\%$}
         \put(24.1,55.65){\boldmath\color{white}$\bullet$}
         \put(24.1,55.65){\boldmath\color{black}$\circ$}
     \end{overpic}
     &
     \begin{overpic}[width=0.33\textwidth]{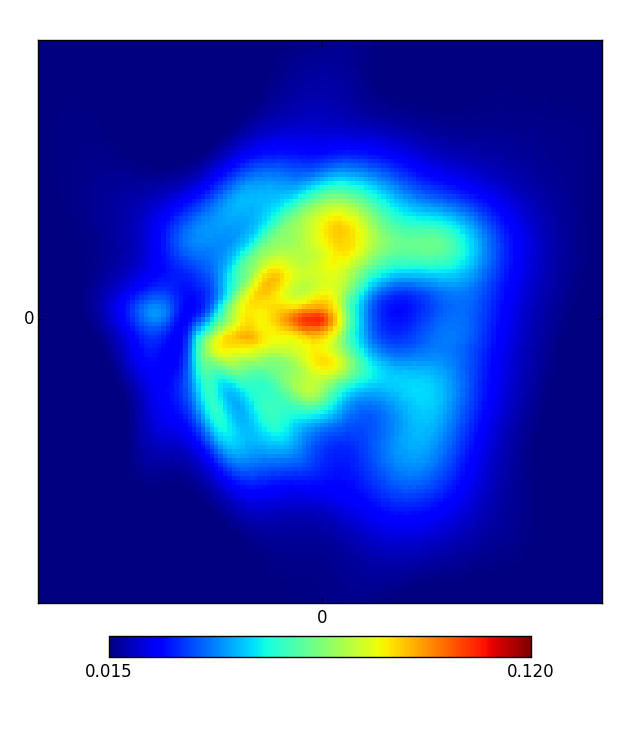}
         \put(30,87){\boldmath\color{white}$10000$ $15\%$}
         \put(24.1,55.65){\boldmath\color{white}$\bullet$}
         \put(24.1,55.65){\boldmath\color{black}$\circ$}
     \end{overpic}
  \\
     \begin{overpic}[width=0.33\textwidth]{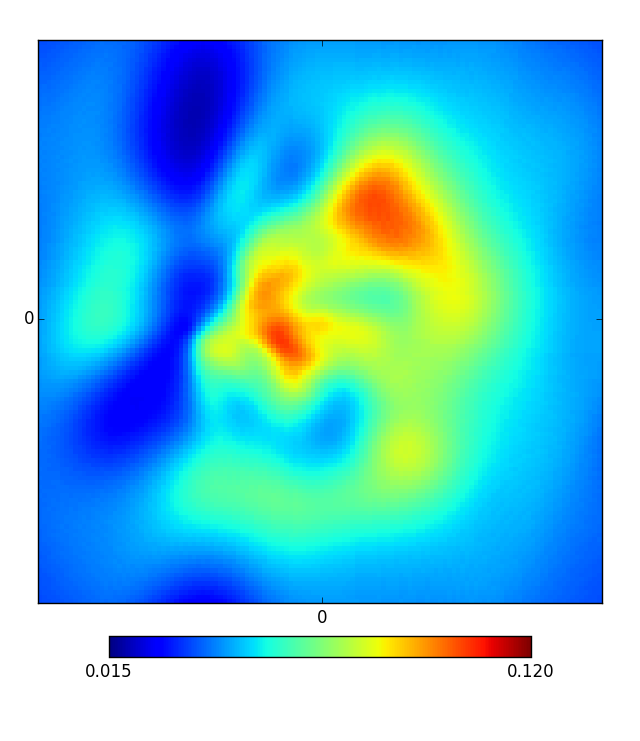}
         \put(35,87){\boldmath\color{white}$1000$ $5\%$}
         \put(24.1,55.65){\boldmath\color{white}$\bullet$}
         \put(24.1,55.65){\boldmath\color{black}$\circ$}
     \end{overpic}
     &
     \begin{overpic}[width=0.33\textwidth]{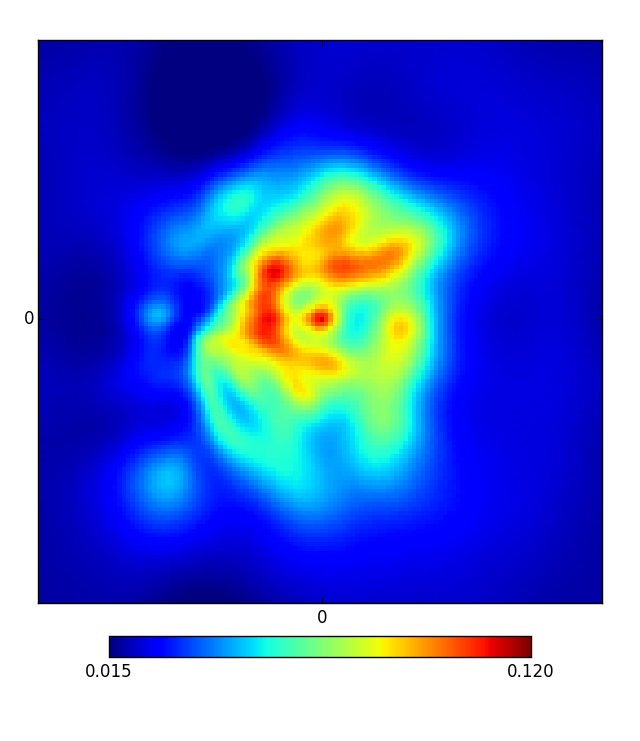}
         \put(35,87){\boldmath\color{white}$5000$ $5\%$}
         \put(24.1,55.65){\boldmath\color{white}$\bullet$}
         \put(24.1,55.65){\boldmath\color{black}$\circ$}
     \end{overpic}
     &
        \begin{overpic}[width=0.33\textwidth]{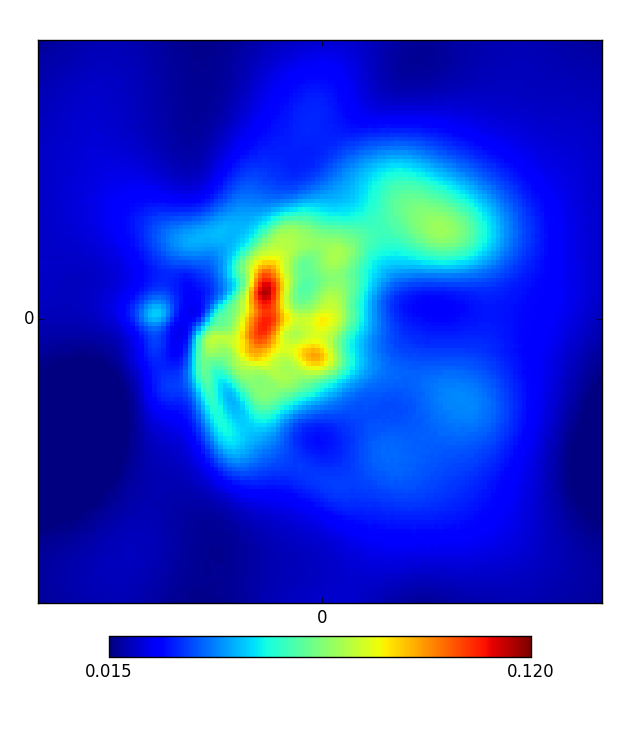}
            \put(30,87){\boldmath\color{white}$10000$\ \textbf{mixed}}
            \put(24.1,55.65){\boldmath\color{white}$\bullet$}
            \put(24.1,55.65){\boldmath\color{black}$\circ$}
        \end{overpic}
   \\
 \end{tabular}
 \caption{Several reconstructions of the Galaxy model with medium strength fluctuations. All panels show top-down views of the \corr{electron} density in the Galactic plane \corr{using a} linear color scale in units of $\mathrm{cm}^{-3}$. \corr{The panels span $36000\,\mathrm{pc}$ in each dimension. The} Sun is located at the white dot depicted in \corr{each} panel.
 The rows show reconstructions with distance errors \corr{of} $25\%$, $15\%$, and $5\%$ respectively (from top to bottom). The columns show reconstructions with $1000$, $5000$, and $10000$ pulsars respectively (from left to right). The layout follows Table~\ref{table:data_sets}. The top left panel shows the original input model (modified NE2001). The bottom right panel shows the reconstruction of the mixed measurement set. 
 }
 \label{fig:compilation}
\end{figure*}

The simulations show that with the amount of pulsars \corr{with} reliable distance estimates \corr{that the SKA should deliver} reconstruction of the free electron density in the vicinity of the Sun becomes feasible (see Fig.~\ref{fig:compilation}). \corr{However, small-scale features are difficult to identify in the reconstruction. Identifying} spiral arms remains challenging \corr{as well}, especially beyond the Galactic center. \corr{To resolve the spiral arms in the vicinity of the Sun, between 5000 and 10000 pulsars with distance accuracies between $5\%$ and $15\%$ are needed.}
As is evident from the figure, \corr{small} distance uncertainties increase the quality of the reconstruction significantly. The reconstruction from 5000 pulsars with $5\%$ distance uncertainty is better in quality than the one from 10000 pulsars with $15\%$ distance uncertainty\footnote{
In principle, this behavior is not surprising, as one measurement of a scalar quantity $a$ with standard deviation $\sigma$ contains the same amount of information as 9 independent measurements with standard deviation $3\sigma$ (assuming Gaussian PDFs).
}.
All reconstructions \corr{smooth out small-scale structure in the electron density, for example at the Galactic center.}
\corr{If an over-density appears at the wrong location this indicates that the data do not constrain the overdensity well.
For completeness we also show the recovered Galactic profile in the Galactic plane for} 5000 pulsars with $5\%$ distance uncertainty \corr{in Fig.~\ref{fig:profile-Gplane}. For other data sets the plot would look very similar.}
\corr{In Appendix~\ref{sec:cheated} we show a reconstruction where the Galactic profile and the correlation structure a known \textit{a priori} and in Appendix~\ref{sec:uncertainty} we show and discuss} the uncertainty estimate of the algorithm.

\begin{figure*}[p]
 \begin{tabular}{c c c}
     \begin{overpic}[width=0.33\textwidth]{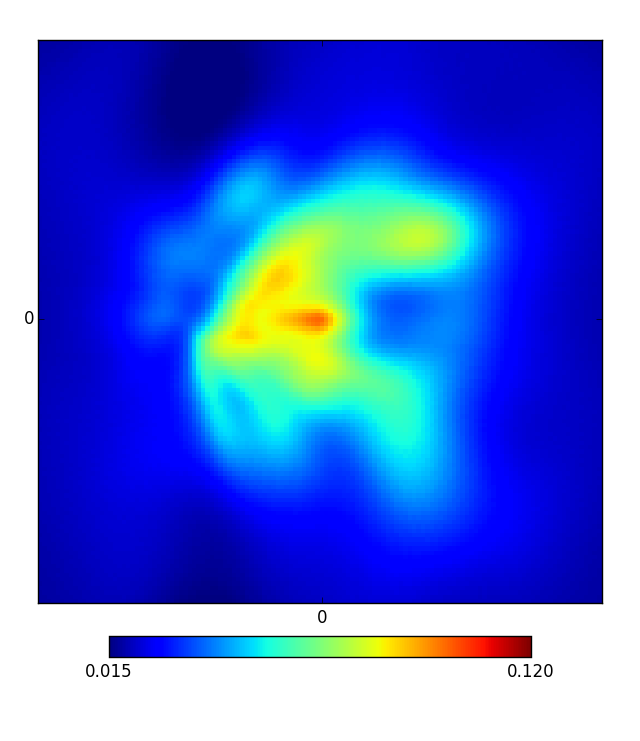}
         \put(33,87){\boldmath\color{white}\textbf{weak} $15\%$}
         \put(24.1,55.65){\boldmath\color{white}$\bullet$}
         \put(24.1,55.65){\boldmath\color{black}$\circ$}
     \end{overpic}
     &
     \begin{overpic}[width=0.33\textwidth]{mid_5000_15}
         \put(30,87){\boldmath\color{white}\textbf{medium} $15\%$}
         \put(24.1,55.65){\boldmath\color{white}$\bullet$}
         \put(24.1,55.65){\boldmath\color{black}$\circ$}
     \end{overpic}
     &
     \begin{overpic}[width=0.33\textwidth]{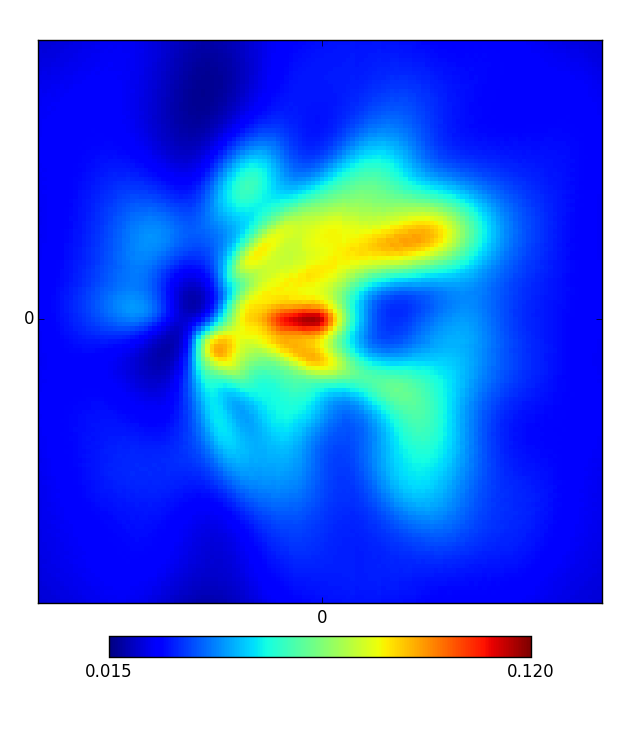}
         \put(31,87){\boldmath\color{white}\textbf{strong} $15\%$}
         \put(24.1,55.65){\boldmath\color{white}$\bullet$}
         \put(24.1,55.65){\boldmath\color{black}$\circ$}
     \end{overpic}
  \\
     \begin{overpic}[width=0.33\textwidth]{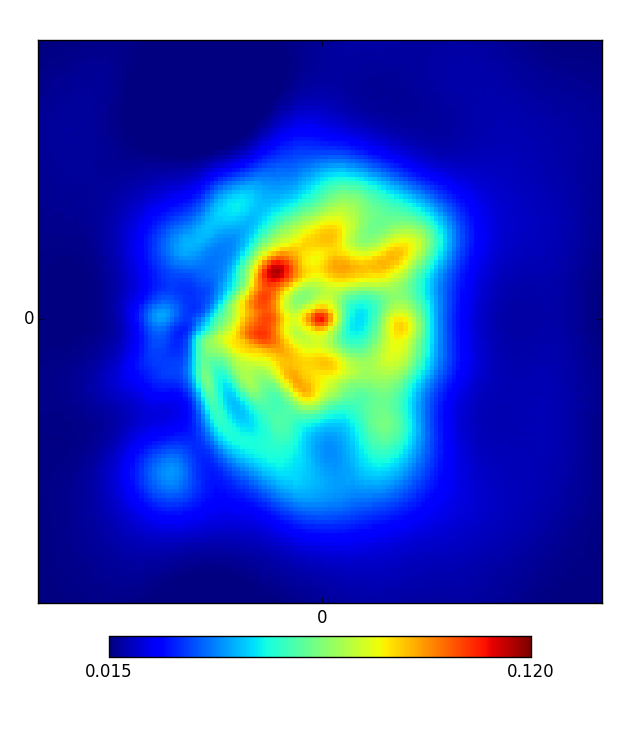}
         \put(35,87){\boldmath\color{white}\textbf{weak} $5\%$}
         \put(24.1,55.65){\boldmath\color{white}$\bullet$}
         \put(24.1,55.65){\boldmath\color{black}$\circ$}
     \end{overpic}
     &
     \begin{overpic}[width=0.33\textwidth]{mid_5000_05}
         \put(32,87){\boldmath\color{white}\textbf{medium} $5\%$}
         \put(24.1,55.65){\boldmath\color{white}$\bullet$}
         \put(24.1,55.65){\boldmath\color{black}$\circ$}
     \end{overpic}
     &
     \begin{overpic}[width=0.33\textwidth]{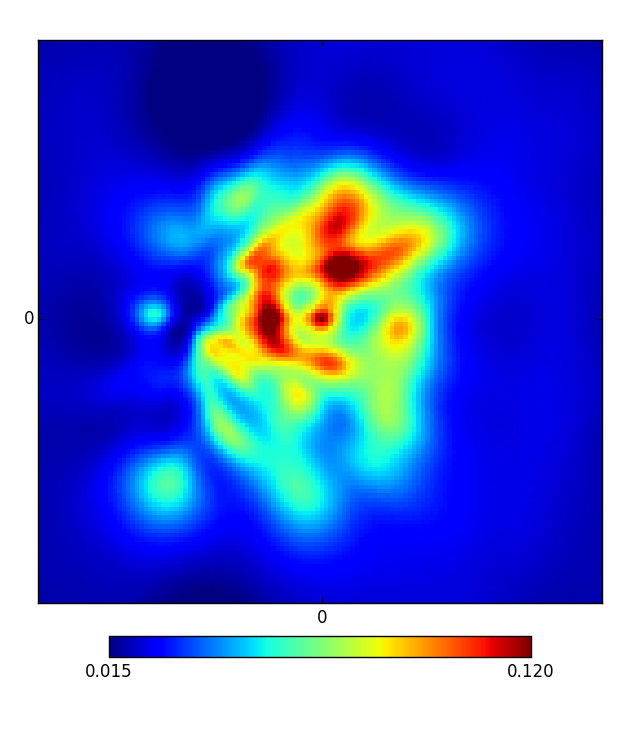}
         \put(33,87){\boldmath\color{white}\textbf{strong} $5\%$}
         \put(24.1,55.65){\boldmath\color{white}$\bullet$}
         \put(24.1,55.65){\boldmath\color{black}$\circ$}
     \end{overpic}
  \\
     \begin{overpic}[width=0.33\textwidth]{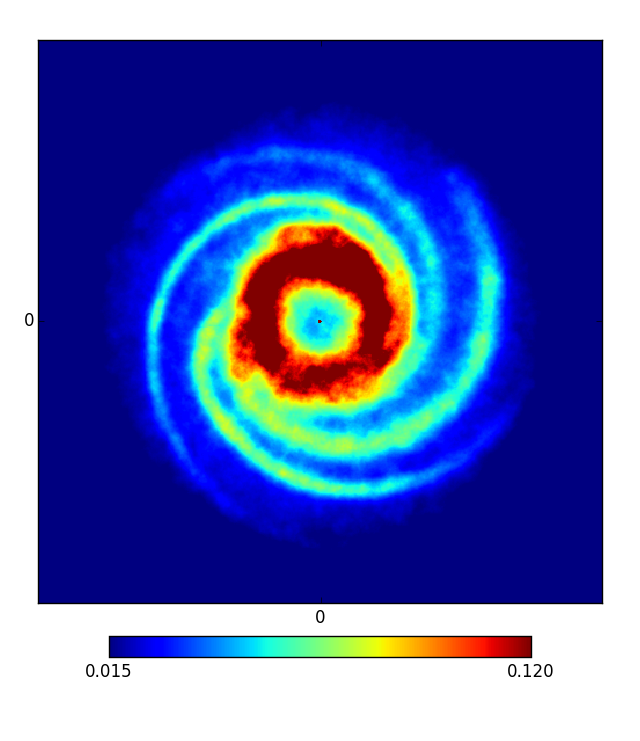}
         \put(28,87){\boldmath\color{white}\textbf{weak original}}
         \put(24.1,55.65){\boldmath\color{white}$\bullet$}
         \put(24.1,55.65){\boldmath\color{black}$\circ$}
     \end{overpic}
     &
     \begin{overpic}[width=0.33\textwidth]{mid_original}
         \put(25,87){\boldmath\color{white}\textbf{medium original}}
         \put(24.1,55.65){\boldmath\color{white}$\bullet$}
         \put(24.1,55.65){\boldmath\color{black}$\circ$}
     \end{overpic}
     &
     \begin{overpic}[width=0.33\textwidth]{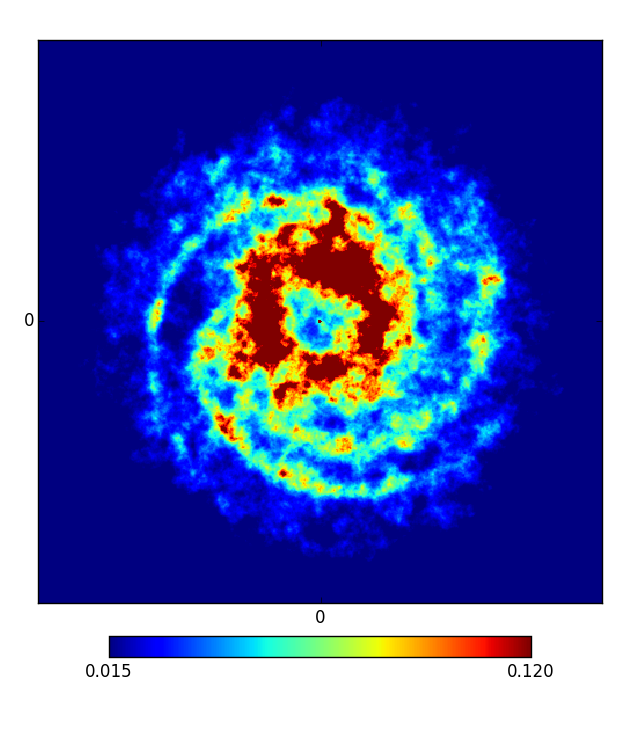}
         \put(26,87){\boldmath\color{white}\textbf{strong original}}
         \put(24.1,55.65){\boldmath\color{white}$\bullet$}
         \put(24.1,55.65){\boldmath\color{black}$\circ$}
     \end{overpic}
   \\
 \end{tabular}
 \caption{Reconstructions of the three Galaxy models with fluctuation strengths using 5000 pulsars. All panels show top-down views of the \corr{electron} density in the Galactic plane \corr{using a} linear color scale in units of $\mathrm{cm}^{-3}$. \corr{The panels span $36000\,\mathrm{pc}$ in each dimension.}
 The rows show reconstructions with distance errors \corr{of} $15\%$ and $5\%$ respectively (from top to bottom). The bottom row shows the original Galaxy models. The columns show reconstructions and original input models (modified NE2001) with weak, medium and strong fluctuations respectively (from left to right). 
 }
 \label{fig:compilation_fluct}
\end{figure*}

In Fig.~\ref{fig:compilation_fluct} we compare the performance of the reconstruction algorithm for the three input model
fluctuation strengths using 5000 pulsars \corr{with} $5\%$ and $15\%$ distance uncertainty. 
One can see that the strength of the fluctuations does not influence the quality of the reconstructions by a great amount. The reconstructions of the models with \corr{stronger fluctuations exhibit stronger} fluctuations as well, while all reconstruction omit/smear features to a similar degree. 
However, one can see that it becomes more difficult to reconstruct the Perseus arm towards the Galactic \corr{anticenter} if the fluctuations in the electron density are strong. This is to be expected, as the spiral arm is also harder to recognize in the original model as the fluctuations become stronger.

\begin{figure}
\centering
 \includegraphics[width=0.33\textwidth]{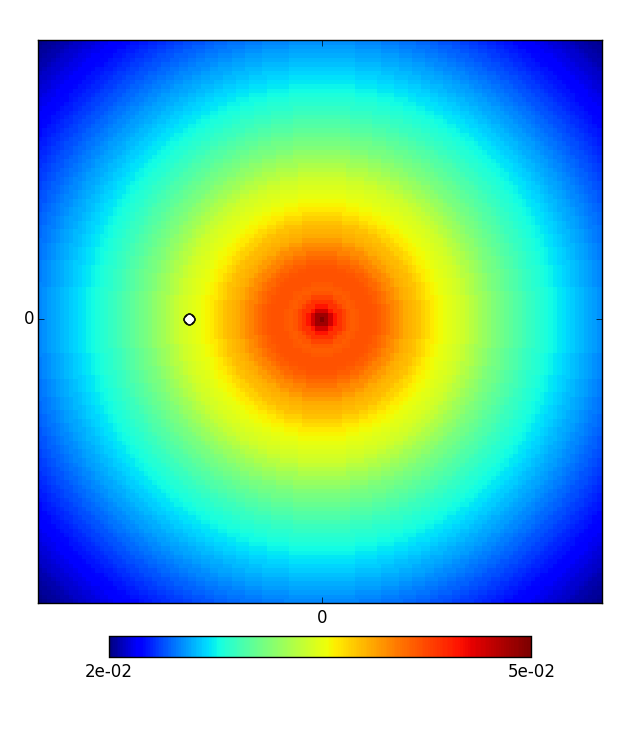}
 \caption{The recovered Galactic profile in the Galactic plane. Shown is a top-down view the Galaxy as in Fig.~\ref{fig:compilation}, but here in logarithmic color scale. The input model had medium strength fluctuations, it was recovered using 5000 pulsars with $5\%$ distance uncertainty (corresponding to the bottom middle panel in Fig.~\ref{fig:compilation}). Other fluctuation strengths and data sets would yield a very similar image.}
 \label{fig:profile-Gplane}
\end{figure}

\begin{figure*}[p]
\centering
 \begin{tabular}{c c}
     \begin{overpic}[width=0.45\textwidth]{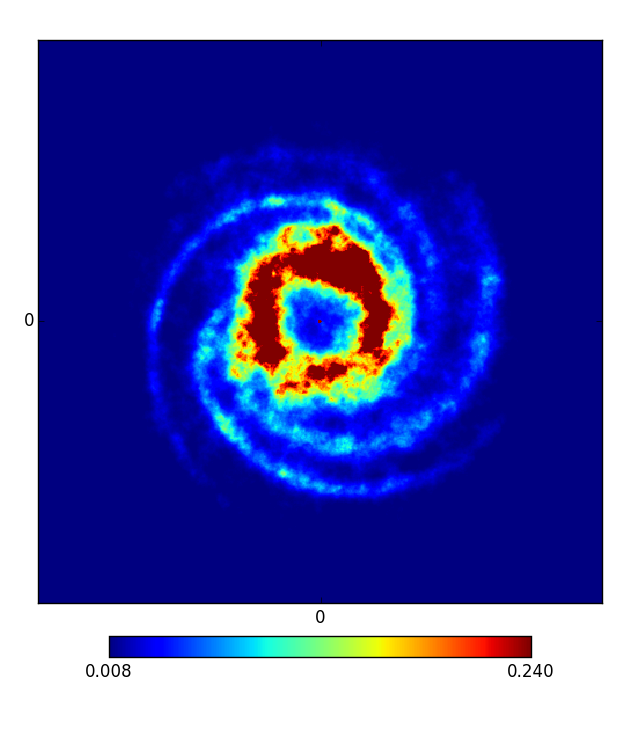}
         \put(29,87){\boldmath\color{white}\textbf{contrast original}}
         \put(24.1,55.65){\boldmath\color{white}$\bullet$}
         \put(24.1,55.65){\boldmath\color{black}$\circ$}
     \end{overpic}
  &
     \begin{overpic}[width=0.45\textwidth]{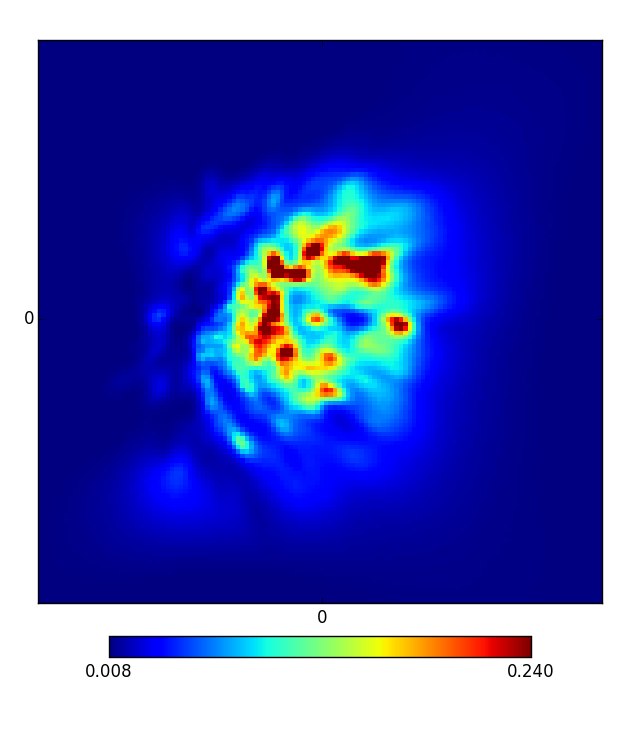}
         \put(30,87){\boldmath\color{white}\textbf{contrast} $5000$ $5\%$}
         \put(24.1,55.65){\boldmath\color{white}$\bullet$}
         \put(24.1,55.65){\boldmath\color{black}$\circ$}
     \end{overpic}
   \\
 \end{tabular}
 \caption{\corr{Input model (left) and reconstruction (right) of the contrast enhanced Galaxy model using 5000 pulsars with distance uncertainties of $5\%$. Both panels show top-down views of the electron density in the Galactic plane using a linear color scale in units of $\mathrm{cm}^{-3}$. The panels span $36000\,\mathrm{pc}$ in each dimension.}
 }
 \label{fig:contrast}
\end{figure*}

\corr{
In Fig.~\ref{fig:contrast} we show the contrast enhanced Galaxy model as well as its reconstruction using 5000 pulsars with distance uncertainties of $5\%$. As is clear from the Figure, the algorithm is able to resolve much more detailed structure compared to the reconstruction of the unenhanced Galaxy model (bottom middle panel in Fig.~\ref{fig:compilation}). We want to stress that the pulsar population and their distance uncertainties are exactly the same for both cases. The increase in quality comes merely from the increased contrast and the resulting stronger imprint of under- and overdensities in the dispersion data. Therefore, we conclude that if the contrast of the real Galaxy is much stronger that in NE2001, our algorithm could resolve the Galaxy much better that the a study on NE2001 indicates.
}

\subsection{Vertical fall-off}
\label{sec:scale-heights}

\begin{figure}
\centering
 \includegraphics[width=0.48\textwidth,trim=0mm 25mm 0mm 80mm,clip=true]{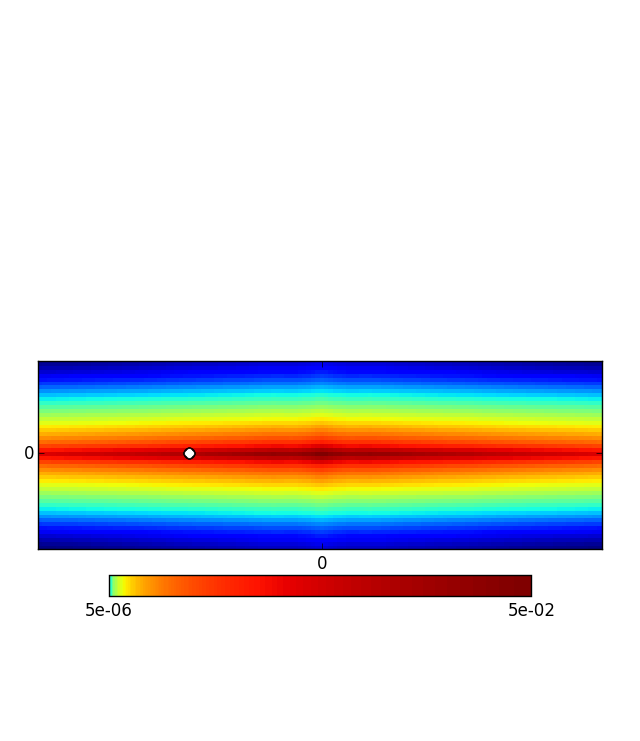}
 \caption{A vertical cut through the same Galactic profile as in Fig.~\ref{fig:profile-Gplane}. Shown is the slice containing the Sun (white dot) and the Galactic center (middle). The image spans $36\,\mathrm{kpc}\times12\,\mathrm{kpc}$. The color scale is logarithmic.}
 \label{fig:profile-vert}
\end{figure}

A quantity of interest in any model of the Galactic free electron density is the drop-off of the average density with respect to distance from the Galactic plane. \corr{This behavior can be seen in Fig.~\ref{fig:profile-vert}, which displays a vertical cut through the Galactic profile reconstructed using 5000 pulsars with $5\%$ distance uncertainty.}

\begin{figure*}[p]
 \begin{tabular}{c c c}
     
        \begin{overpic}[width=0.33\textwidth,trim=20mm 0mm 15mm 5mm,clip=true]{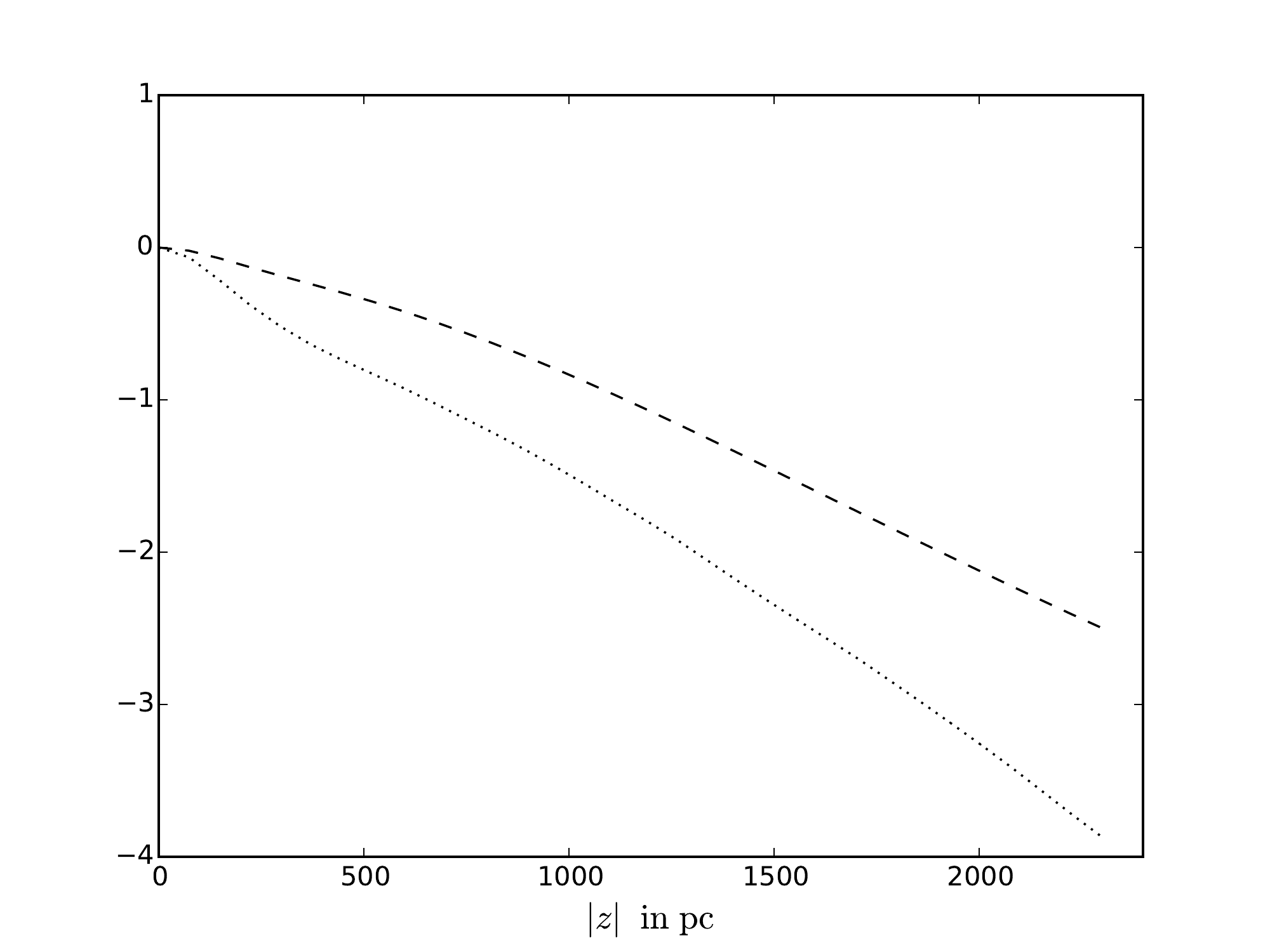}
            \put(40,70){\boldmath\color{black}\textbf{original}}
            \put(9,22){\color{black}$H_z=610\,\mathrm{pc}$ (local)}
            \put(9,15){\color{black}$H_z=880\,\mathrm{pc}$ (global)}
        \end{overpic}
     
     &
     \begin{overpic}[width=0.33\textwidth,trim=20mm 0mm 15mm 5mm,clip=true]{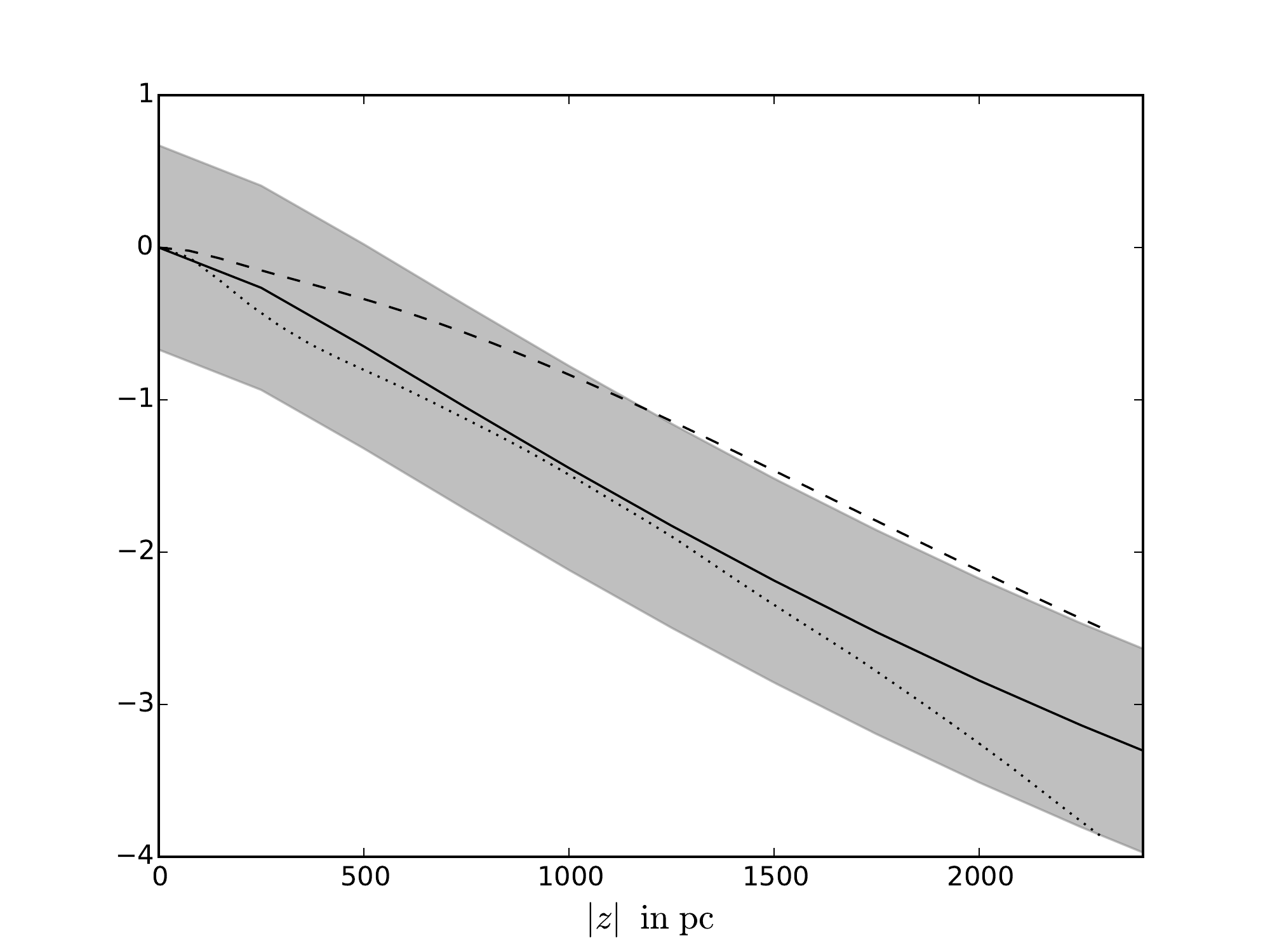}
         \put(40,70){\boldmath\color{black}$5000$ $25\%$}
         \put(9,15){\color{black}$H_z=(740\pm120)\,\mathrm{pc}$}
     \end{overpic}
     &
     \begin{overpic}[width=0.33\textwidth,trim=20mm 0mm 15mm 5mm,clip=true]{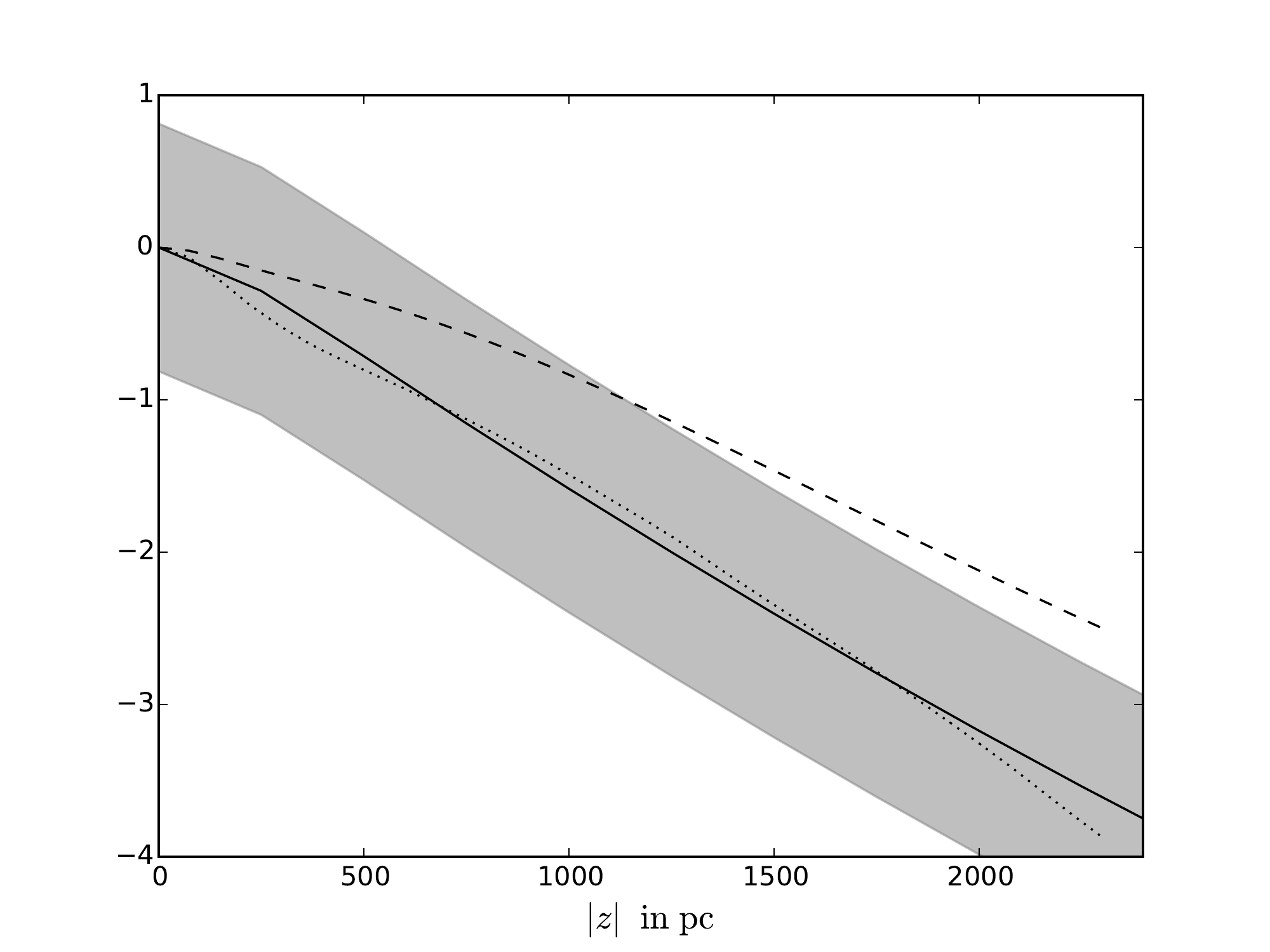}
         \put(40,70){\boldmath\color{black}$10000$ $25\%$}
         \put(9,15){\color{black}$H_z=(630\pm80)\,\mathrm{pc}$}
     \end{overpic}
  \\
     \begin{overpic}[width=0.33\textwidth,trim=20mm 0mm 15mm 5mm,clip=true]{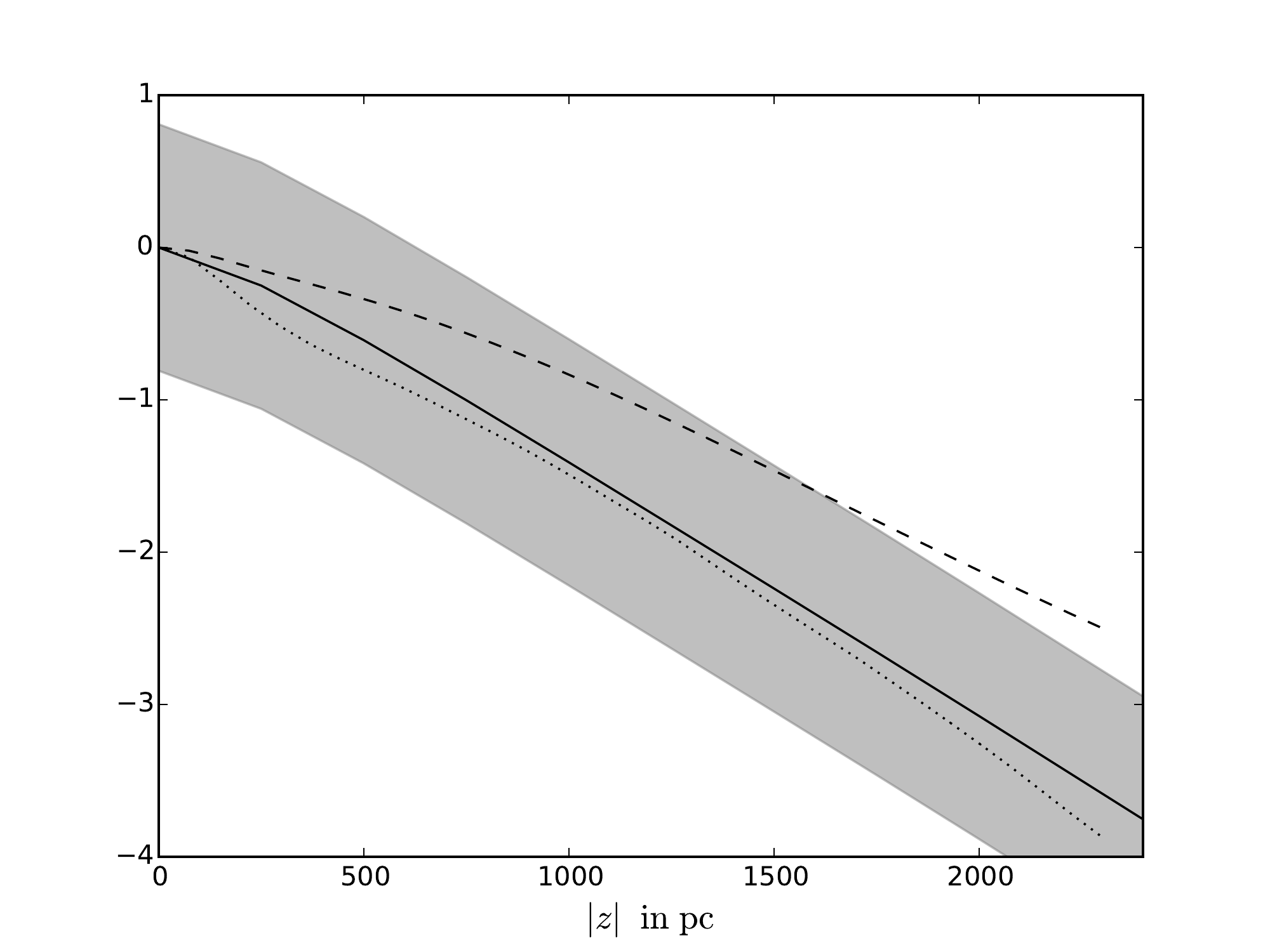}
         \put(40,70){\boldmath\color{black}$1000$ $15\%$}
         \put(9,15){\color{black}$H_z=(650\pm110)\,\mathrm{pc}$}
     \end{overpic}
     &
     \begin{overpic}[width=0.33\textwidth,trim=20mm 0mm 15mm 5mm,clip=true]{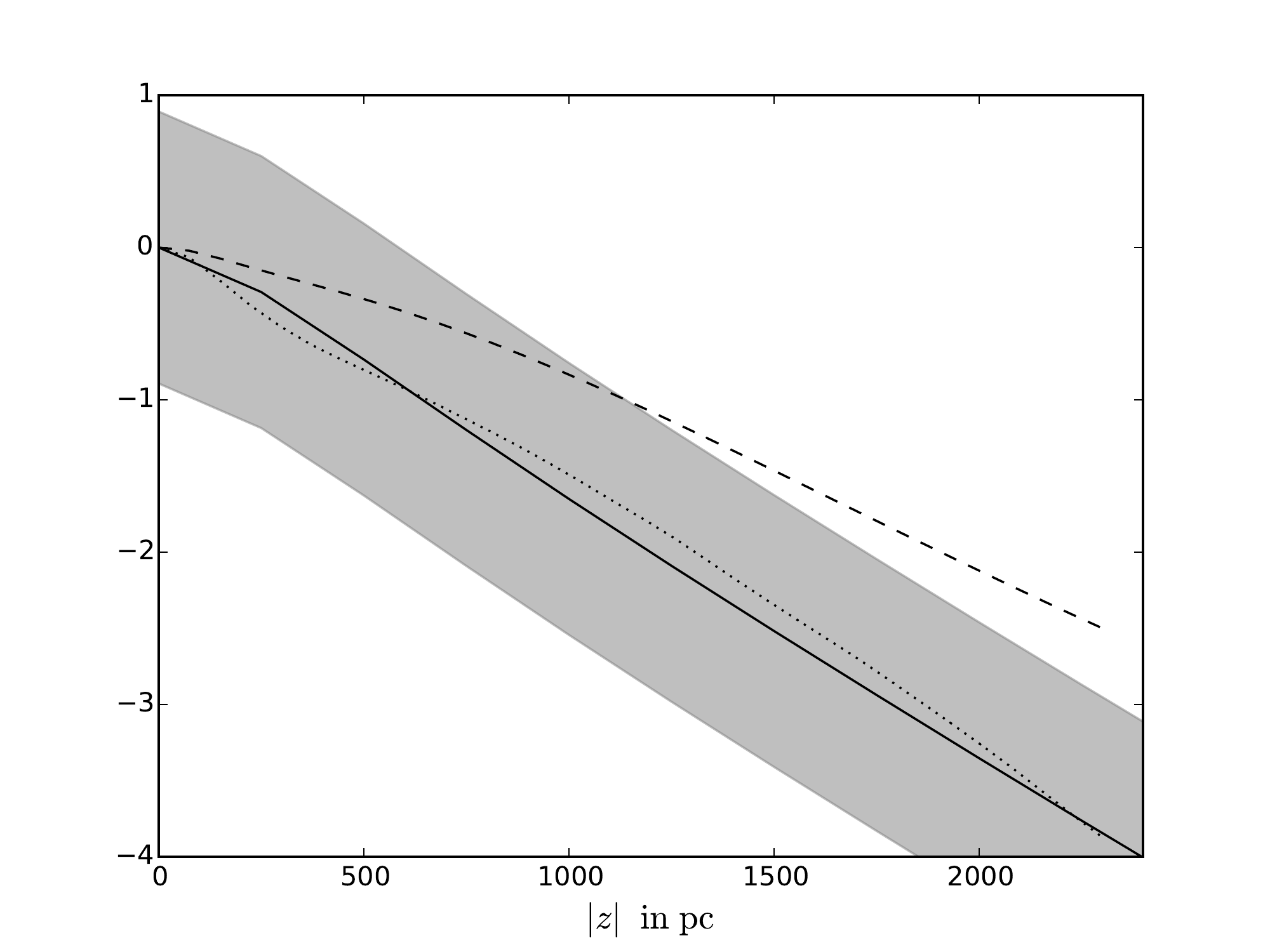}
         \put(40,70){\boldmath\color{black}$5000$ $15\%$}
         \put(9,15){\color{black}$H_z=(590\pm70)\,\mathrm{pc}$}
     \end{overpic}
     &
     \begin{overpic}[width=0.33\textwidth,trim=20mm 0mm 15mm 5mm,clip=true]{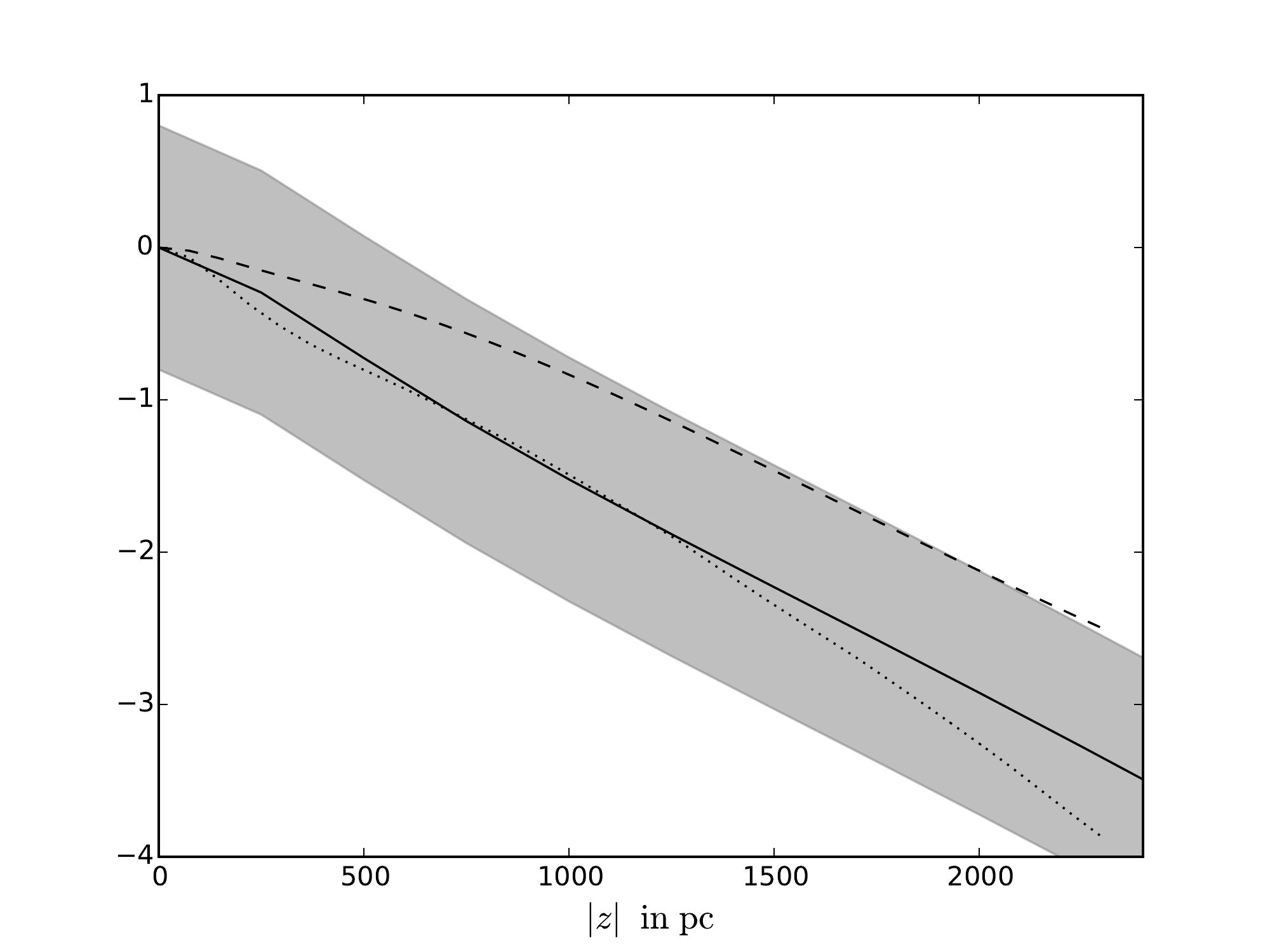}
         \put(40,70){\boldmath\color{black}$10000$ $15\%$}
         \put(9,15){\color{black}$H_z=(690\pm120)\,\mathrm{pc}$}
     \end{overpic}
  \\
     \begin{overpic}[width=0.33\textwidth,trim=20mm 0mm 15mm 5mm,clip=true]{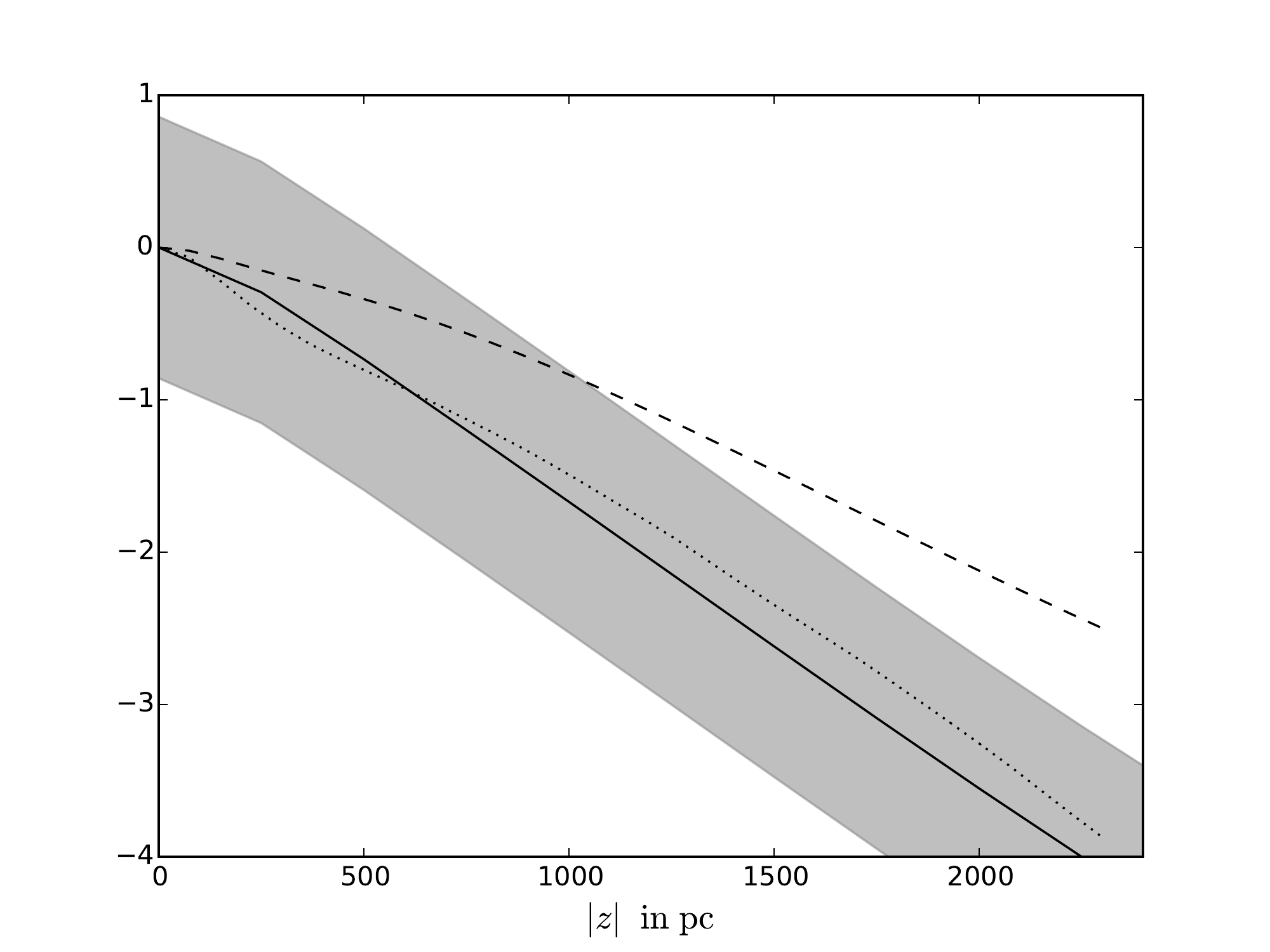}
         \put(40,70){\boldmath\color{black}$1000$ $5\%$}
         \put(9,15){\color{black}$H_z=(550\pm60)\,\mathrm{pc}$}
     \end{overpic}
     &
     \begin{overpic}[width=0.33\textwidth,trim=20mm 0mm 15mm 5mm,clip=true]{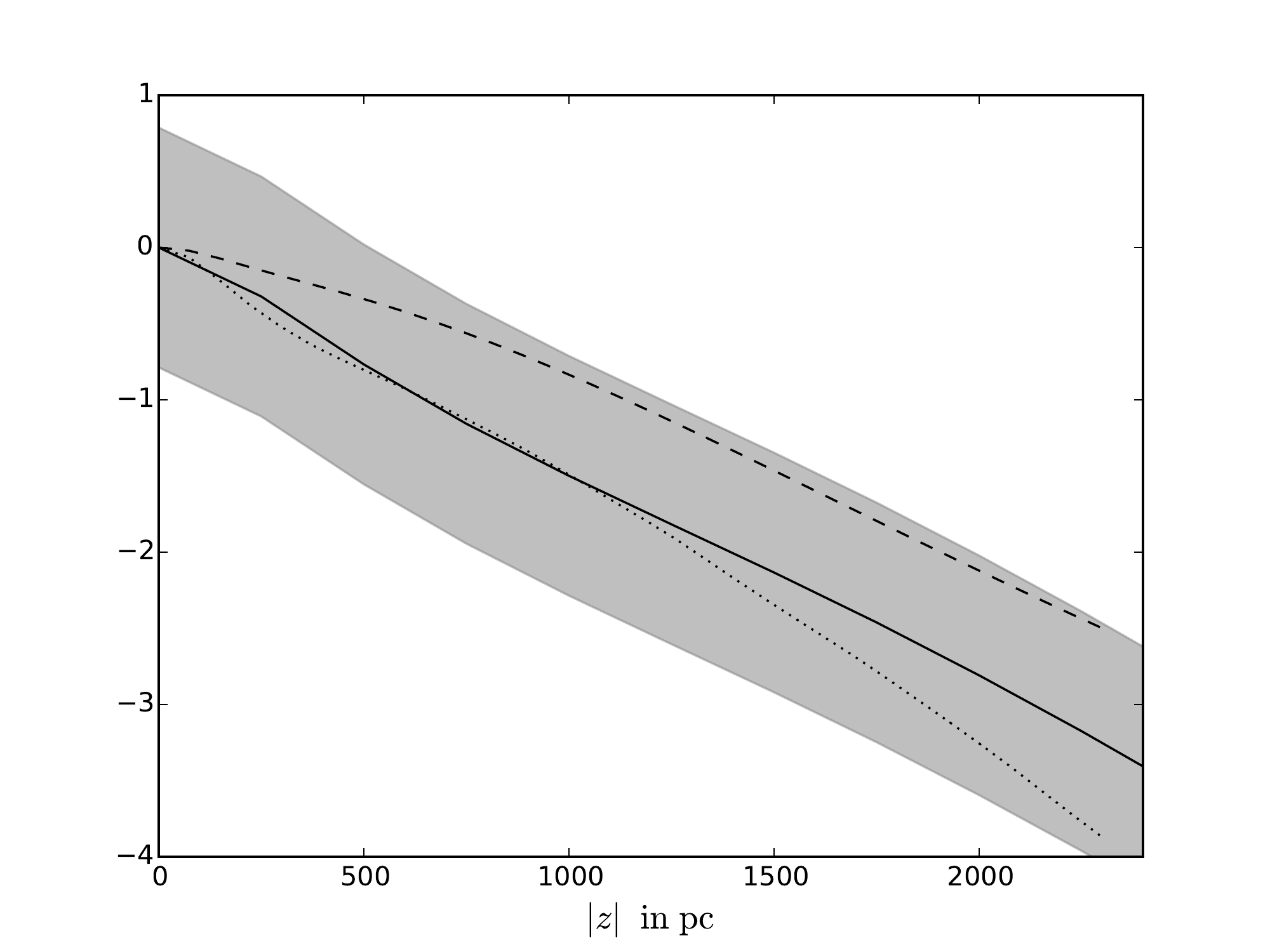}
         \put(40,70){\boldmath\color{black}$5000$ $5\%$}
         \put(9,15){\color{black}$H_z=(740\pm130)\,\mathrm{pc}$}
     \end{overpic}
     &
        \begin{overpic}[width=0.33\textwidth,trim=20mm 0mm 15mm 5mm,clip=true]{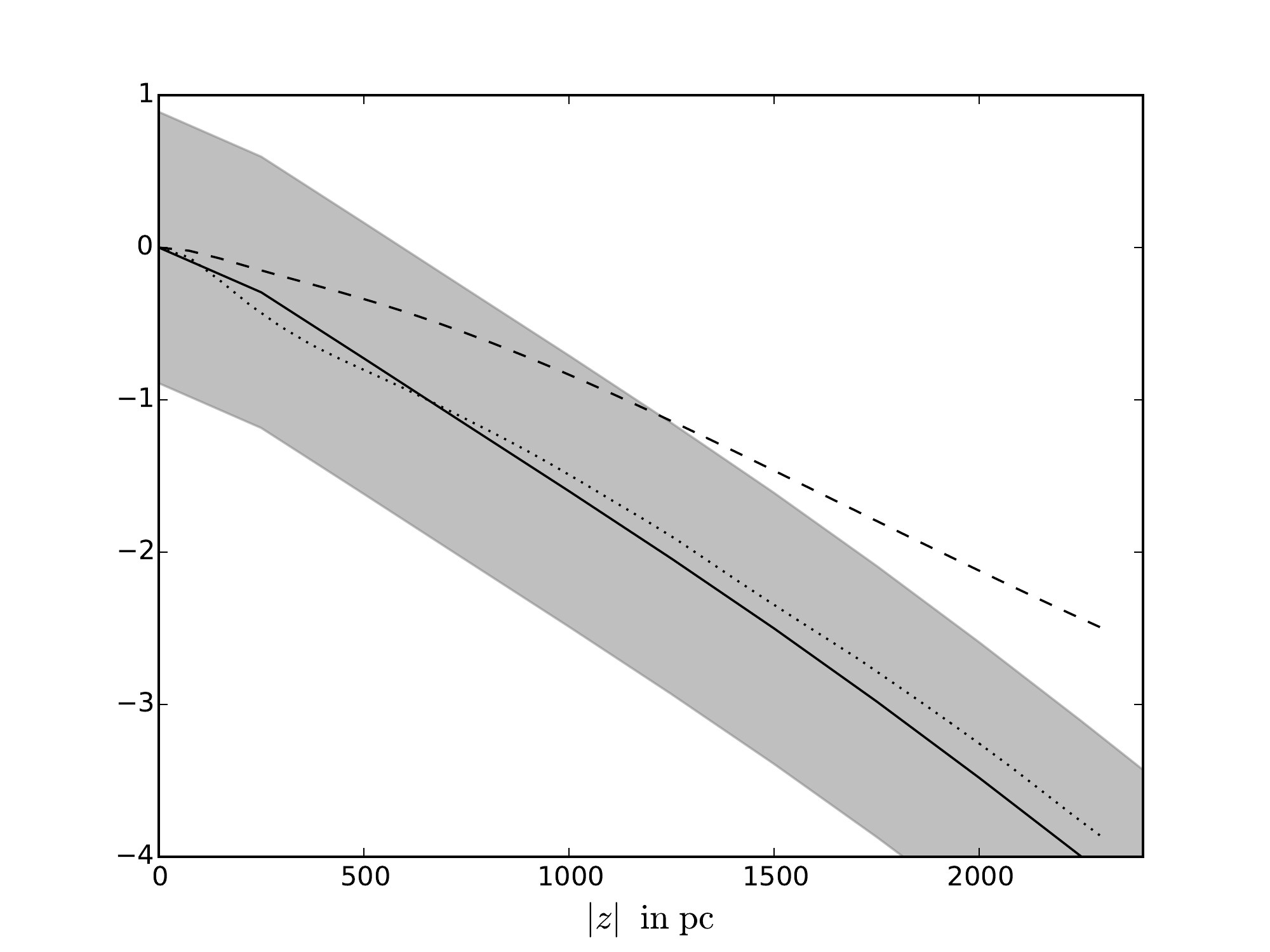}
            \put(35,70){\boldmath\color{black}$10000$\ \textbf{mixed}}
            \put(9,15){\color{black}$H_z=(570\pm80)\,\mathrm{pc}$}
        \end{overpic}
     
   \\
 \end{tabular}
 \caption{The recovered z-dependent fall-off in logarithmic units (function $\beta$ in Eq.~\eqref{eq:profiles}), input model with medium strength fluctuations. The top left panel shows the global z-profile (dashed line) as well as the local z-profile (dotted line). In all other panels the solid line is the recovered z-profile while the global and local z-profile are replotted (in dashed and dotted respectively). The gray areas indicate the $1\sigma$ uncertainty around the recovered z-profile. In the bottom left corner of each panel we show the best fitting exponential scale height (and its 1$\sigma$ uncertainty for the reconstructions).}
 \label{fig:compilation_zprof}
\end{figure*}

\corr{In} our parametrization the function $\beta$ in Eq.~\eqref{eq:profiles} describes the average log-density at a certain distance from the Galactic plane. In Fig.~\ref{fig:compilation_zprof} we show the estimates for $\beta$ corresponding to the reconstructions shown in Fig.~\ref{fig:compilation} along with their uncertainties (see Appendix~\ref{sec:z_profile_uncertainity} for their calculation). The uncertainty regions reflect that a vertical fall-off can be explained by a global profile as well as by density fluctuations \corr{close to the Sun. This uncertainty is nearly independent of the quality of the data set, but depends on the strength of fluctuations on kpc scales. These are always present unless the data probe a simplistic disk. Therefore, there is a lower bound of precision to which our algorithm can determine the vertical fall-off behavior.}
\corr{We compare the reconstructed vertical scaling to a global} and a local estimate \corr{generated from the original input model.} The global estimate describes vertical fall-off throughout the whole model whereas the local estimate describes the vertical fall-off close to the Sun\footnote{ 
The global estimate is calculated by averaging the logarithmic density at fixed vertical distances over the whole horizontal plane. The local estimate is calculated by averaging the logarithmic density at fixed vertical distances in a sub-area of the horizontal plane, which is centered on Sun and has a size of $1500\,\mathrm{pc}\times1500\,\mathrm{pc}$.
}. \corr{For completeness we also provide best fitting scale heights for exponential fall-offs in the figure, i.e., we fit the vertical scaling to $n_\mathrm{e} \propto e^{-|z|/H_z}$. The uncertainties of these estimates are calculated by performing the fit on multiple posterior samples of $\beta$. Both, the local and the global estimate have significantly lower scale heights than the $950\,\mathrm{pc}$ from NE2001 (thick disk). This is probably due to the combination of the thick disk with the thin disk of NE2001 (which has a scale height of $140\,\mathrm{pc}$).}
As is evident from the figure, the reconstructed z-profile is dominated by the local behavior of the density and agrees with it within the error bars throughout all data sets\footnote{ 
The reconstructed vertical fall-off is dominated by the near-Sun region since this is the part of the Galaxy where the density is reconstructed best.
}, for the regime $|z| < 2400\,\mathrm{pc}$. \corr{However, the width of the uncertainty region prohibits a clear decision whether the vertical fall-off follows a single exponential function or a thick disk and a thin disk, as is the case for NE2001. 
In our input model we set $n_\mathrm{e}$ to zero for $|z|>2400\,\mathrm{pc}$. In that regime our reconstruction is unreliable.}

\section{Summary and conclusions}
\label{sec:discussion}

We presented an algorithm that performs \corr{nonparametric tomography of the} Galactic free electron density using pulsar dispersion measures and distances. The algorithm produces a three-dimensional map and a corresponding uncertainty map. It estimates the correlation structure and the scales of the disk shape automatically, requiring only approximate initial guesses for them. The uncertainties of pulsar distance estimates are consistently propagated. 

Using our algorithm we investigated the feasibility of \corr{nonparametric} tomography with the upcoming Square Kilometer Array. To that end, we created three Galaxy models \corr{with various} fluctuation strengths \corr{and one with enhanced contrast} and \corr{simulated mock observations of these models} using between 1000 and 10000 pulsars. Our results indicate that with the amount of pulsars \corr{that the SKA should deliver, nonparametric tomography becomes feasible.} However, detecting spiral arms in the free electron density from pulsar dispersion measures alone remains challenging \corr{if the input model has unenhanced contrast}. We find that to distinguish the spiral arms in the vicinity of the Sun, between 5000 and 10000 pulsars with distance accuracies between $5\%$ and $15\%$ are needed. The vertical fall-off behavior of the free electron density was recovered for all mock data sets we investigated. However, a clear decision whether the vertical fall-off of free electron density is best described by a single exponential function or a thick disk and a thin disk \corr{could not be made by} our algorithm.

\corr{One way to increase the sensitivity of the algorithm for Galactic features would be to include them into the prior description. Including higher order statistics (non-Gaussian priors) one could make the inference more sensitive for spiral arm structures in the electron density. In cosmology, for example, higher order statistics allowed for a better recovery of cosmological filaments (e.g. \cite{Jasche-2013}). Modeling of HII regions and supernova remnants is also beyond the scope of Gaussian statistics.
Another approach would be to include parametrized structures which are known from stellar observations, e.g., spatial templates for spiral arm locations. This would connect data driven tomography (with infinite degrees of freedom) with classical model fitting.}

The algorithm can also be used for other tomography problems with line-of-sight measurements, such as stellar absorption coefficients. It could also be extended to infer vector fields, enabling inference of the Galactic magnetic fields from pulsar rotation measures. Furthermore, a joint reconstruction of the Galactic free electron density and the magnetic field using pulsar dispersion, measures, pulsar rotation measures as well as extragalactic Faraday sources should be investigated.

\begin{acknowledgements}
 We want to thank Niels Oppermann and Marco Selig for fruitful collaboration and advice. We also want to thank Henrik Junklewitz, Jimi Green, Jens Jasche and Sebastian Dorn for useful discussions. The calculations were realized using the \textsc{NIFTY}\footnote{\url{http://www.mpa-garching.mpg.de/ift/nifty/}} package by \cite{nifty-2013}. Some of the minimizations described in Sec.~\ref{sec:filter_equations} were performed using the \mbox{L-BFGS-B} algorithm (\cite{LBFGS-1995}). We acknowledge the support by the DFG Cluster of Excellence "Origin and Structure of the Universe". The simulations have been carried out on the computing facilities of the Computational Center for Particle and Astrophysics (C2PAP). We are grateful for the support by Frederik Beaujean through C2PAP.
 This research has been partly supported by the DFG Research Unit 1254 and has made use of NASA's Astrophysics Data System.
\end{acknowledgements}

\begin{appendix}

\section{Parameters of the power spectrum prior}
\label{app:parameters}

The inverse Gamma distribution is defined as
\begin{equation}
 \mathcal{I}(p_k;\alpha_k,q_k) = \frac{1}{q_k\Gamma(\alpha_k-1)} \left( \frac{p_k}{q_k} \right)^{-\alpha_k}\, \exp\!\left( - \frac{q_k}{p_k} \right).
\end{equation}
The mean and variance of this distribution are 
\begin{equation}
\begin{split}
 \left\langle p_k \right\rangle_{(p_k)} & = q_k/(\alpha_k-2) \ \  && \mathrm{for}\ \ \alpha>2\\
 \left\langle p_k^2 \right\rangle_{(p_k)} -\left\langle p_k \right\rangle_{(p_k)}^2 & = \frac{q_k^2}{(\alpha_k-3)(\alpha_k-2)^2}  \ \  && \mathrm{for}\ \ \alpha>3.\\
 \end{split}
\end{equation}
There are three properties of the prior that we want to fulfill by choosing $\alpha_k$ and $q_k$. The prior of the monopole $p_0$, which corresponds to the variance of a global prefactor of the density, should be close to Jeffreys prior, i.e. the limit $\alpha_0 \rightarrow 1$, $q_0 \rightarrow 0$. The reason for this is that we do want the algorithm to stay consistent if one changes the units, say go from $\mathrm{pc}$ to $\mathrm{kpc}$. Such changes introduce a global prefactor in front of the density. Jeffreys prior has no preferred scale as it is flat for $\log p_0$ and therefore all prefactors are equally likely \textit{a priori}.
For other $k$-bins the parameters should favor, but not enforce falling power spectra. Furthermore, since $p(k)$ is the average power of many independent Fourier components, its \textit{a priori} variance should be inversely proportional to $\varrho_k$ (the amount of degrees of freedom in the respective $k$-bin), while the \textit{a priori} mean should be independent of $\varrho_k$.
We therefore set the parameters as
\begin{equation}
 \begin{split}
  q_k = f \,\varrho_k \qquad \mathrm{and} \qquad
  \alpha_k = 1 + \frac{k}{100 k_\mathrm{min}} \varrho_k,
 \end{split}
\end{equation}
where $k_\mathrm{min}$ is the first non-zero $k$-value and $f$ is a prefactor, which defines a lower cut-off of the power spectrum calculated in Eq.~\eqref{eq:powspec_approx}. The choice of $f$ does not influence the result as long as it is suitably low, but higher $f$ accelerate the convergence of the algorithm. The denominator of $100 k_\mathrm{min}$ before $\varrho_k$ is chosen to introduce a preference for falling power spectra starting two orders of magnitude from the fundamental mode $k_\mathrm{min}$ (note that the 1 is subtracted in Eq.~\eqref{eq:smoothness-prior}). As long as the denominator is not too small it has very little influence on the result of the algorithm but smaller denominators increase the convergence speed. We found $100 k_\mathrm{min}$ to be a good compromise.

The parameter $\sigma_p$ in Eq.~\eqref{eq:smoothness-prior} describes how much the power spectrum is expected to deviate from a power law. We choose $\sigma_p = 1$. If the power spectrum is locally described by a power law, $\sigma_p = 1$ means that the typical change of the exponent within a factor of $e$ in $k$ should be of order $1$.

\section{Functional derivatives of the Hamiltonian}
\label{app:derivatives}

To minimize the Hamiltonian in Sec.~\ref{sec:filter_equations} the first derivative with respect to $s$ is needed. It is
\begin{equation}
\begin{split}
 \frac{\delta}{\delta s^\dagger} \mathcal{H}(s,D\!M|p,\tilde{N}) & =  S^{-1}s + \widehat{\left(\mathrm{e}^s \right)} M \left(\mathrm{e}^s \right) - \widehat{\left(\mathrm{e}^s \right)} j,
\end{split}
\end{equation}
where the hat converts a field to a diagonal operator in position space, e.g.\ \mbox{$\widehat{\xi}(\vec{x},\vec{y}) = \xi(\vec{x})\delta(\vec{x}-\vec{y})$}, and we used the shorthand notations
\begin{equation}
 \begin{split}
  \qquad S^{-1} & \equiv \sum\limits_k S^{(k)} p_k^{-1}\\
 M & \equiv \widehat{\Delta}\tilde{R}^\dagger \tilde{N}^{-1}\tilde{R}\widehat{\Delta}\\
j & \equiv \widehat{\Delta}\tilde{R}^\dagger \tilde{N}^{-1} D\!M.
 \end{split}
\end{equation}

The second derivative in Eq.~\eqref{eq:cov_estimate} is
\begin{equation}
 \begin{split}
  \frac{\delta^2}{\delta s \delta s^\dagger} \mathcal{H}(s,D\!M|p,\tilde{N}) & = S^{-1} + \widehat{\left(\mathrm{e}^s \right)} M \widehat{\left(\mathrm{e}^s \right)} \\
 & \quad+ \widehat{\left(\mathrm{e}^s \right)} \widehat{M\left(\mathrm{e}^s \right)} - \widehat{\left(\mathrm{e}^s \right)}\,\widehat{j}.
 \end{split}
\end{equation}
The last term in the second derivative can be problematic as it can break the positive definiteness\footnote{Mathematically, the second derivative has to be positive definite at the minimum, but in high dimensional parameter spaces this is not guaranteed in numerical practice.} of the second derivative, which is crucial to apply inversion techniques such as the conjugate gradient method efficiently.
However, a closer inspection of the last two terms (omitting the hats for readability),
\begin{equation}
 M\left(\mathrm{e}^s \right) - j =  \widehat{\Delta}\tilde{R}^\dagger \tilde{N}^{-1} \left( \tilde{R}\widehat{\Delta}\mathrm{e}^s - D\!M \right) \propto \widetilde{D\!M} - D\!M,
\end{equation}
shows that their contribution is proportional to the difference between the real dispersion data $D\!M$ and the idealized data generated by the map, $\widetilde{D\!M}=\tilde{R}\widehat{\Delta}\mathrm{e}^s$. These two terms counteract each other at the minimum and we therefore omit them to gain numerical stability. Hence the second derivative is approximated as
\begin{equation}
 \frac{\delta^2}{\delta s \delta s^\dagger} \mathcal{H}(s,D\!M|p,\tilde{N}) \approx S^{-1} + \widehat{\left(\mathrm{e}^s \right)} M \widehat{\left(\mathrm{e}^s \right)}.
\end{equation}

\section{Convergence}
\label{sec:convergence}

In this section, we display the convergence behavior of the power spectrum, effective errors, and profile functions. For the sake of brevity, we will limit the discussion to the reconstruction of the Galaxy model with fluctuations of medium strength using the data set of 5000 pulsars with a distance error of $5\%$. In Appendix~\ref{sec:cheated} we show the reconstruction of this model and data set using the real power spectrum and profile functions as a benchmark on how well our iterative estimation of them does.

\begin{figure}
 \includegraphics[width=0.5\textwidth]{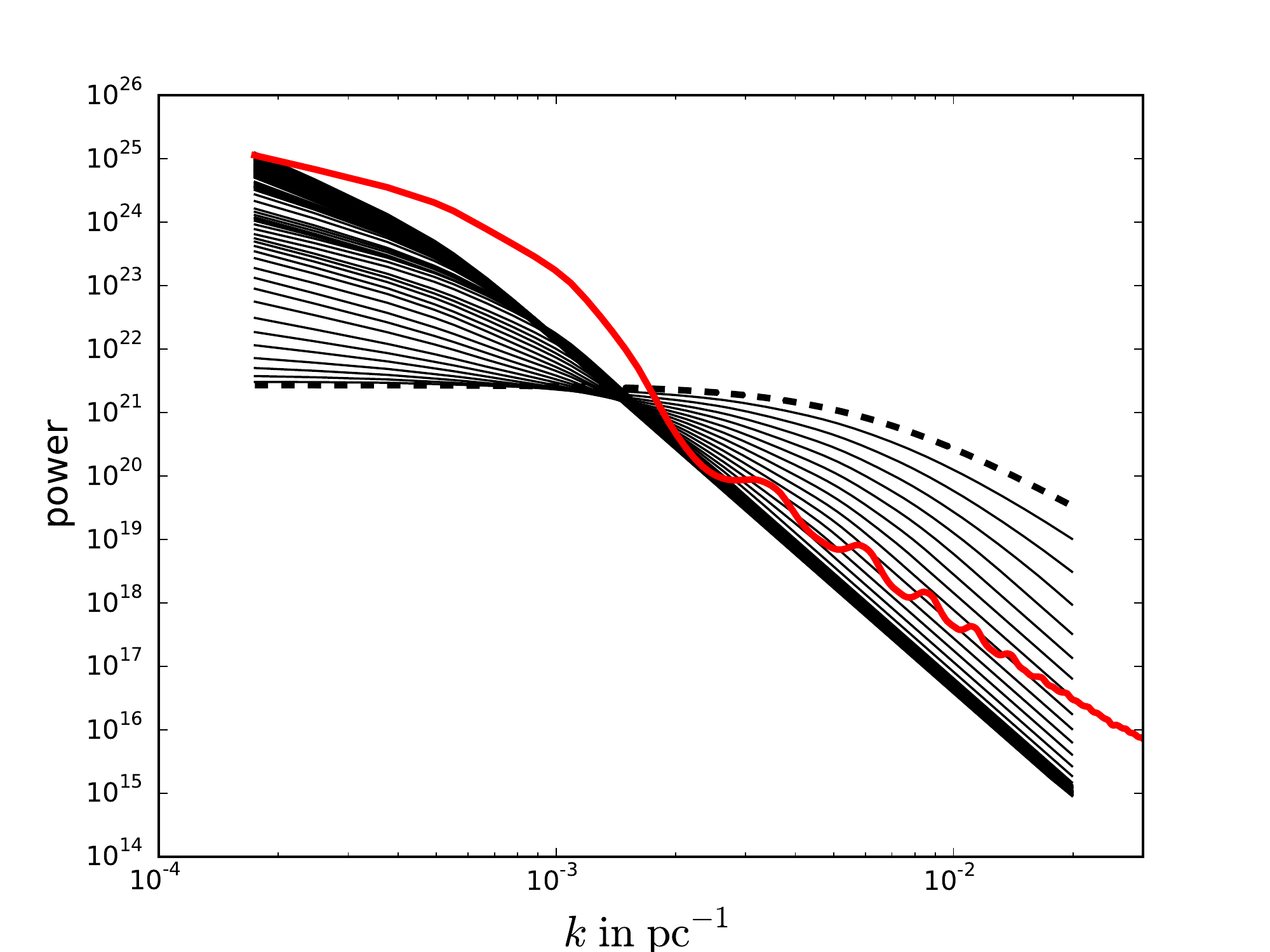}
 \caption{A plot of the power spectrum changing with the iterations. The thick dashed line is the initial guess, the bulge of black lines is where the algorithm converges. The thick solid red line is the power spectrum of the logarithmic Galaxy model with medium fluctuations. The power spectrum is in arbitrary units.}
 \label{fig:power_convergence}
\end{figure}

In Fig.~\ref{fig:power_convergence} we show the convergence of the power spectrum. As is evident the power moves away from the initial guess to a fixed point. Compared to the spectrum of the logarithmic input model the converged spectrum misses power in both, the large-scale and the small-scale regime. The loss of power in the large-scale regime is due to the profile field absorbing large features, in the small-scale regime it is due to the general loss of small-scale power.

The loss of small-scale power comes from two effects. First, the dispersion measure data sample the density sparse and irregularly. Without the regularization imposed by the prior, this would lead to severe aliasing as is commonly known from Fourier analysis. As the prior typically suppresses aliasing from large scales to small scales and the algorithm consequently interpretes the power as noise. Aliasing from small scales to large scales is negligible, since the input model is spatially correlated and has thus a falling power spectrum. Second, there is the loss of power due to the distance uncertainties. These make the the likelihood less informative about small-scale structures which are in consequence surpressed by the prior. This effect yields no aliasing but smoothens the resulting map (which is desired to avoid overfitting). For further details about the loss of power in filtering algorithms such as the one in this paper we refer to \cite{Crit-2011}.

The fixed point power spectrum falls as a power law with index $-5.5$ for $k>2\times10^{-4}$. \corr{Our algorithm allows for spectral indices up to $-5.5$.} Without this limit, the power spectrum would fall\footnote{
This means that the algorithm underestimates the power on scales which are not sufficiently probes by the data set. This does not influence the quality of the density map too much, but it makes the algorithm underestimate the posterior uncertainty.
} to a minimal value of $q_k/\varrho_k$ for $k\gtrsim 3\times 10^{-4}$. However, introducing a hard limit speeds up the convergence and using a slope of $5.5$ made no difference towards the lower limit for the resulting maps in our tests. One can think of this hard limit to be part of the power spectrum prior.

\begin{figure}
 \includegraphics[width=0.5\textwidth]{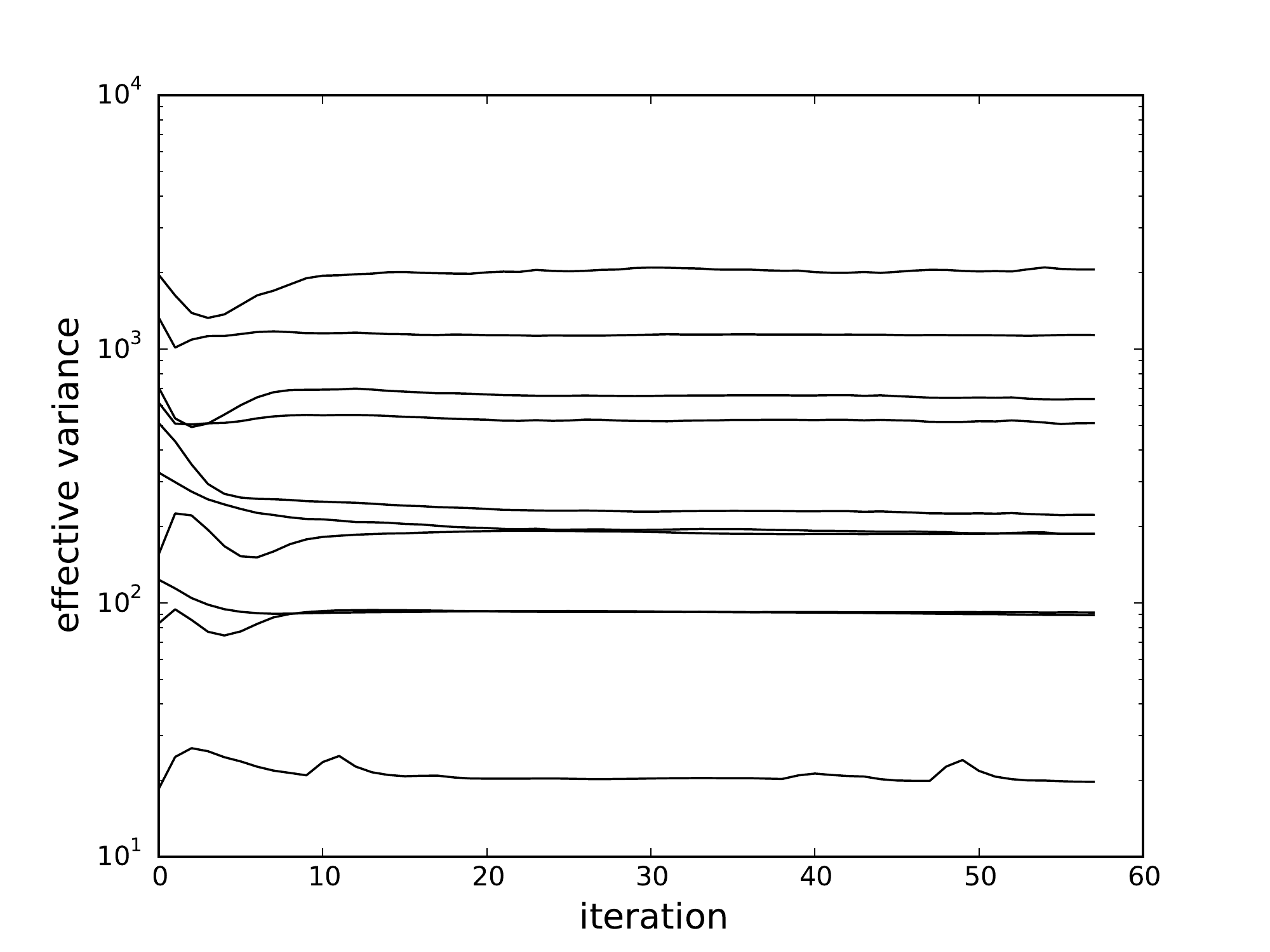}
 \caption{The propagated distance variance of 10 data points changing with the iterations. The units of the variances are $\left(\frac{\mathrm{pc}}{\mathrm{cm}^3}\right)^2$.}
 \label{fig:error_convergence}
\end{figure}

In Fig.~\ref{fig:error_convergence} we show the convergence of the propagated distance variance (Eq.~\ref{eq:noise_addition}) of a random selection of 10 data points. The behavior seen in this plot is qualitatively the same for all data points we investigated. As one can see, most data points reach convergence rather quickly, but there are also outliers. In this plot, the lowest line exhibits a kink after it had seemingly already converged. Such behavior is unfortunately not entirely suppressible, but it appears to have very little effect on the resulting map, as only a small fraction of data points does this.

\begin{figure}
 \includegraphics[width=0.5\textwidth]{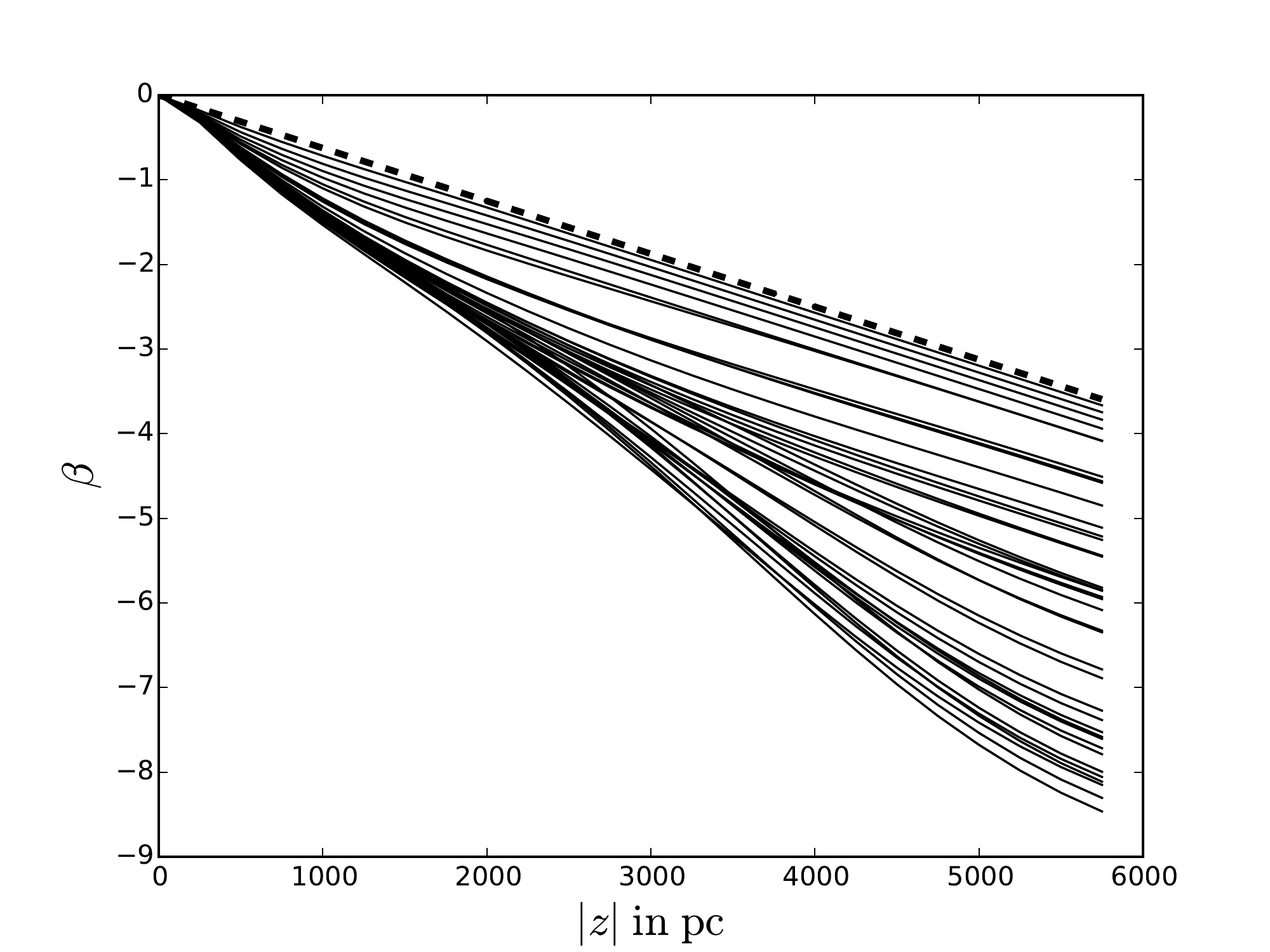}
 \caption{The z-profile function of of $\log n_\mathrm{e}$ changing with the iterations. The thick dashed line is the initial guess.} \label{fig:zfunc_convergence}
\end{figure}

\begin{figure}
 \includegraphics[width=0.5\textwidth]{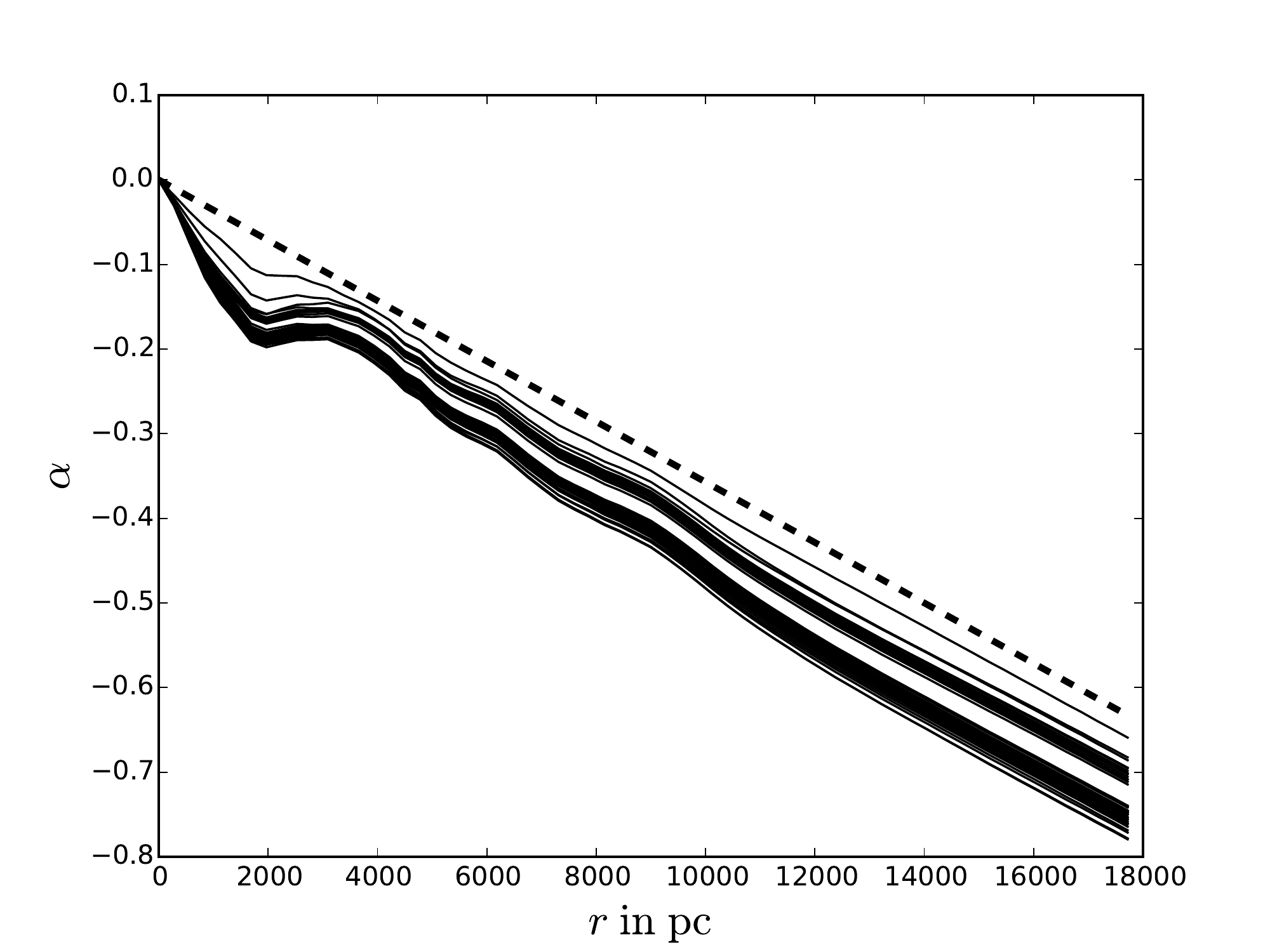}
 \caption{The radial profile function of $\log n_\mathrm{e}$ changing with the iterations. The thick dashed line is the initial guess.}
 \label{fig:rfunc_convergence}
\end{figure}

In Figs.~\ref{fig:zfunc_convergence}~and~\ref{fig:rfunc_convergence} we show the convergence of the profile functions, where we shifted the functions by a global value to line them up at $\beta(|z|\!=\!0)$ and $\alpha(r\!=\!0)$. We note that the functions $\alpha$ and $\beta$ are degenerate with respect to a global addition in their effect on the Galactic profile field and degenerate with the monopole of $s$ as well. This is why a shift by a constant for plotting purposes is reasonable. The z-profile function $\beta$ seems to reach a fixed point for $|z|<2400\,\mathrm{pc}$. For higher $|z|$ the profile function reaches no clear fixed point. However, for the Galactic profile, where the profile function is exponentiated, this makes only a small difference, since $\beta$ is already three $\mathrm{e}$-foldings below its values at $|z|=0$. The radial profile function $\alpha$ seems to only correct the initial guess mildly and it is not clear whether the result is independent from the initial guess. However, it appears that 
$\alpha$ does reach a fixed point.

\section{Goodness of fit ($\chi^2$) of the reconstructions}
\label{sec:chisquared}

In this section, we discuss the goodness of fit characterized by the reduced $\chi^2$ value,
\begin{equation}
 \chi^2 = \frac{1}{N_\mathrm{data}}\sum\limits_{i=1}^{N_\mathrm{data}} \frac{\left(D\!M_i - \widetilde{D\!M}_i\right)^2}{\sigma^2_i},
\end{equation}
where $\widetilde{D\!M}$ is the dispersion data reproduced by applying the response to our reconstruction,
\begin{equation}
 \widetilde{D\!M}_i = \left(\tilde{R}m^{(\rho)}\right)_i,
\end{equation}
and $\sigma_i$ is the crudely propagated distance uncertainty given by 
\begin{equation}
 \sigma_i = \frac{\sqrt{\mathrm{Var}[d_i]}}{d_i} D\!M_i.
\end{equation}
\corr{To put the $\chi^2$ value into perspective we compare them with the $\chi^2$ values of the null model ($m^{(\rho)} = 0$) and the Galactic profile only ($m^{(\rho)} = \Delta$)}
The reduced $\chi^2$ values corresponding to the maps shown in Fig.~\ref{fig:compilation} are shown in Table~\ref{tab:compilation_chi}. The reconstruction of the mixed data set has a reduced $\chi^2$ of 1.31.
In Table~\ref{tab:compilation2_chi}, we show the reduced $\chi^2$ values of the maps shown in Fig.~\ref{fig:compilation_fluct}. \corr{For the reconstruction of the contrast enhanced input model shown in Fig.~\ref{fig:contrast} the reduced $\chi^2$ value is $345$ for the null model, $110$ for the profile only and $2.6$ for the full reconstrution.}

\begin{table}
 \caption{The reduced $\chi^2$ values corresponding to the maps shown in Fig.~\ref{fig:compilation}.}
 \label{tab:compilation_chi}
 \begin{tabular}{ r c c c }
  \hline\hline
  data set & null model & profile only & full map\\
  \hline
  5000 PSR @ 25\% & 17.0 & 1.53 & 0.89 \\
  10000 PSR @ 25\% & 16.9 & 1.48 & 0.89 \\
  1000 PSR @ 15\% & 44.4 & 3.20 & 1.21 \\
  5000 PSR @ 15\% & 44.7 & 3.78 & 1.24 \\
  10000 PSR @ 15\% & 44.6 & 3.48 & 1.11 \\
  1000 PSR @ \phantom{1}5\% & 345 & 23.2 & 3.34 \\
  5000 PSR @ \phantom{1}5\% & 345 & 23.5 & 2.56 \\
 \end{tabular}
\end{table}

\begin{table}
 \caption{The $\chi^2$ values corresponding to the maps shown in Fig.~\ref{fig:compilation_fluct}.}
 \label{tab:compilation2_chi}
 \begin{tabular}{ r c c c }
  \hline\hline
  data set & null model & profile only & full map\\
  \hline
   weak @ 15\% & 44.7 & 2.67 & 1.05 \\
   medium @ 15\% & 44.7 & 3.78 & 1.24 \\
   strong @ 15\% & 44.7 & 6.44 & 1.54 \\
   weak @ \phantom{1}5\% & 345 & 16.2 & 2.34 \\
   medium @ \phantom{1}5\% & 345 & 23.5 & 2.56 \\
   strong @ \phantom{1}5\% & 345 & 47.1 & 3.01 \\
 \end{tabular}
\end{table}
\corr{It is evident from the tables, that our reconstructions account for a large fraction of the data variance in all cases. The Galactic profile without local fluctuations also accounts for a large fraction of the variance, especially if the distance uncertainties are high and the flucutaion strength of the input model is weak.
For our reconstructions} the $\chi^2$ values are close to 1 for the $25\%$ and $15\%$ data sets. Therefore, we assume that our inference mechanism resolved the most relevant information in the data sets and that the prior assumptions are not too restrictive for these data sets. For the $5\%$ reconstructions the $\chi^2$ values are around $3$. This is a hint that the data might contain more information than the reconstruction resolves and that more elaborate prior assumptions might yield a better map. However, how to achieve this is a non-trivial question and we do not aim to answer it in this work.

\section{Reconstruction with real power spectrum and profile functions}
\label{sec:cheated}

\begin{figure*}
  \begin{tabular}{c c c}
     \begin{overpic}[width=0.33\textwidth]{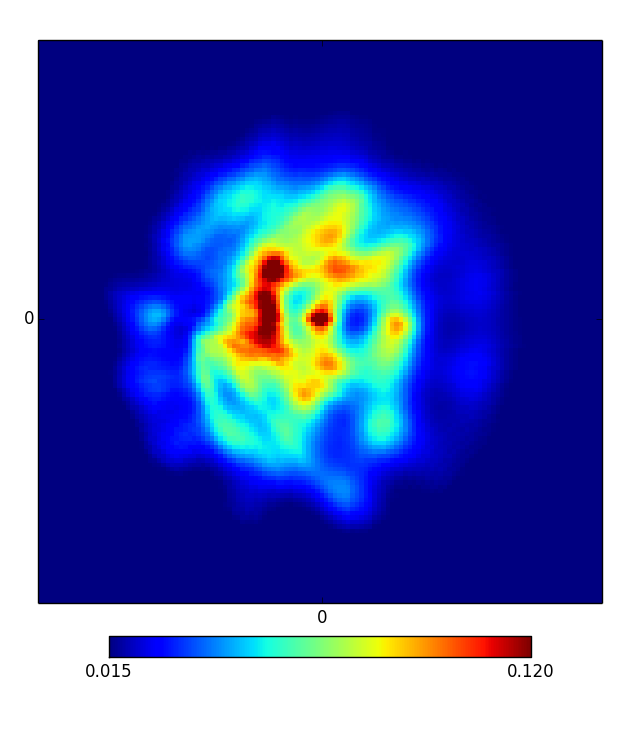}
         \put(34,87){\boldmath\color{white}\textbf{cheated}}
         \put(24.1,55.65){\boldmath\color{white}$\bullet$}
         \put(24.1,55.65){\boldmath\color{black}$\circ$}
     \end{overpic}
     &
     \begin{overpic}[width=0.33\textwidth]{mid_5000_05}
         \put(34,87){\boldmath\color{white}\textbf{inferred}}
         \put(24.1,55.65){\boldmath\color{white}$\bullet$}
         \put(24.1,55.65){\boldmath\color{black}$\circ$}
     \end{overpic}
     &
     \begin{overpic}[width=0.33\textwidth]{mid_original}
         \put(34,87){\boldmath\color{white}\textbf{original}}
         \put(24.1,55.65){\boldmath\color{white}$\bullet$}
         \put(24.1,55.65){\boldmath\color{black}$\circ$}
     \end{overpic}
  \\
  \end{tabular}
 \caption{Top-down view of the reconstructed electron densities in the Galactic plane (in units of $\mathrm{cm}^{-3}$), if we use the real power spectrum and Galactic profile functions (``cheated''), the results from our algorithm (``inferred''), and the original input model.}
 \label{fig:cheated}
\end{figure*}

The posterior map our algorithm finds depends on the prior power spectrum, the effective errors, and the profile functions, all of which are simultaneously estimated from the data. To benchmark the efficiency of this joint estimation, we investigate the case where the real power spectrum as well as the real profile functions are known, i.e.~only iterating the effective errors. The resulting map serves as an indicator whether our Ansatz with unknown hyper parameters is sensible or whether the problem is too unrestricted in that setting.
We depict the map resulting from the real hyper parameters in Fig.~\ref{fig:cheated}. As one can see, the morphology of the result does not change. \corr{More of the small-scale structure is resolved} and the intensity of the overdensity between Sun and the Galactic center, which belongs to the ring the original model, is more pronounced. Consequently, this map has a better reduced $\chi^2$ value of $1.95$ compared to the value of $2.56$ of our map with unknown hyper parameters. But considering the amount of unknowns this is a satisfactory result. We therefore regard our estimation procedure for the hyper parameters as sensible.

\section{Uncertainty map}
\label{sec:uncertainty}

\begin{figure*}
  \begin{tabular}{c c}
     \begin{overpic}[width=0.33\textwidth]{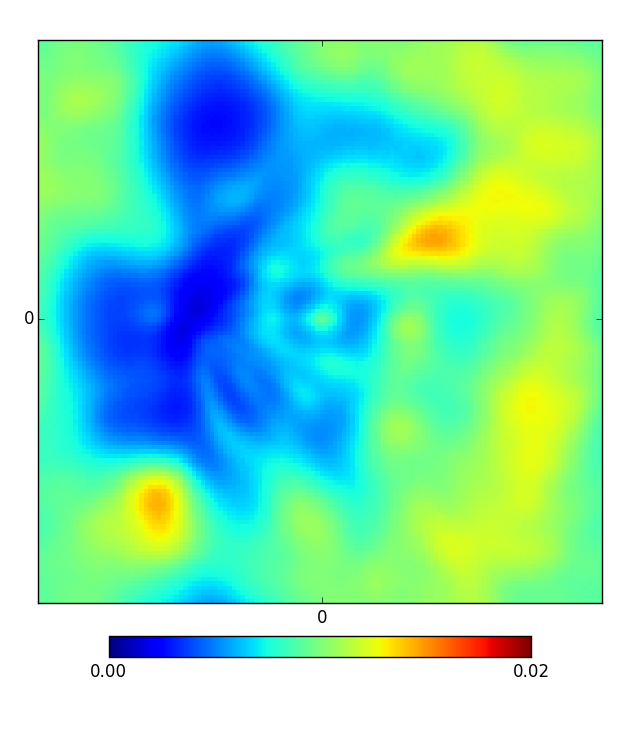}
         \put(35,97){\boldmath\color{black}{\textbf{inferred}}}
         \put(40,87){\boldmath\color{black}\large{\textbf{$\sigma^{(\rho)}$}}}
         \put(24.1,55.65){\boldmath\color{white}$\bullet$}
         \put(24.1,55.65){\boldmath\color{black}$\circ$}
     \end{overpic}
     &
     \begin{overpic}[width=0.33\textwidth]{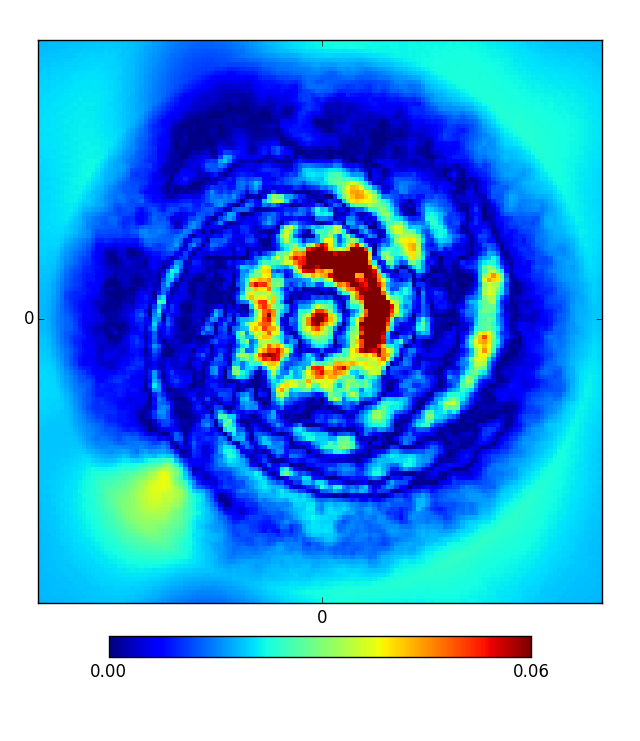}
         \put(35,97){\boldmath\color{black}{\textbf{inferred}}}
         \put(32,87){\boldmath\color{white}\large{\textbf{$|m^{(\rho)}-\rho|$}}}
         \put(24.1,55.65){\boldmath\color{white}$\bullet$}
         \put(24.1,55.65){\boldmath\color{black}$\circ$}
     \end{overpic}
  \\
     \begin{overpic}[width=0.33\textwidth]{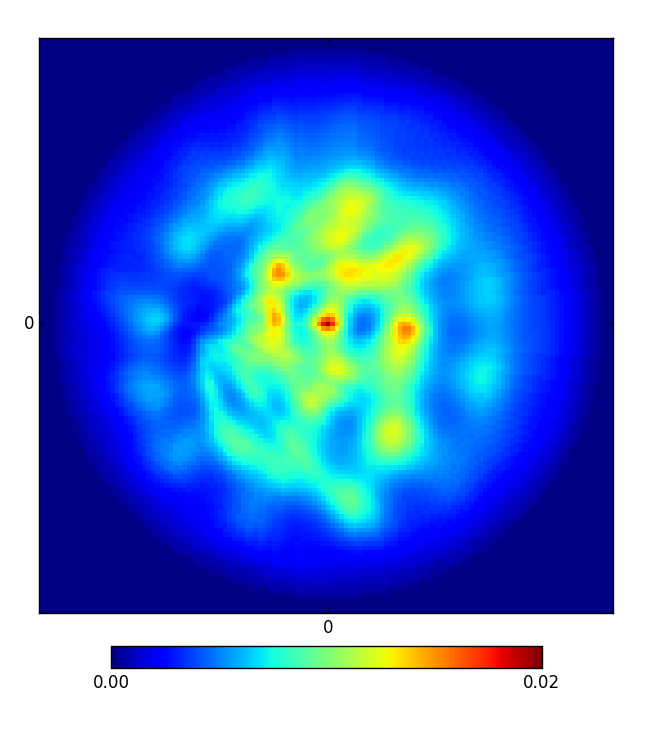}
         \put(35,97){\boldmath\color{black}{\textbf{cheated}}}
         \put(40,87){\boldmath\color{white}\large{\textbf{$\sigma^{(\rho)}$}}}
         \put(24.1,55.65){\boldmath\color{white}$\bullet$}
         \put(24.1,55.65){\boldmath\color{black}$\circ$}
     \end{overpic}
     &
     \begin{overpic}[width=0.33\textwidth]{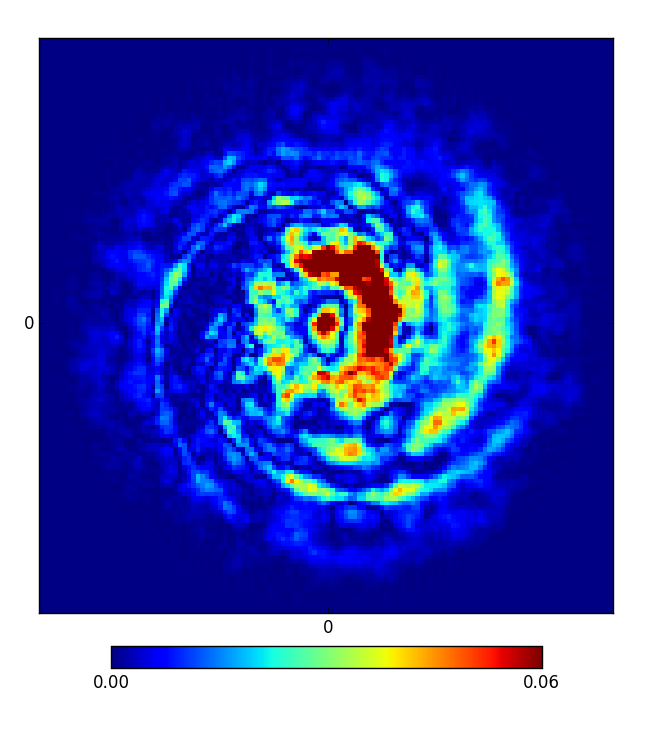}
         \put(35,97){\boldmath\color{black}{\textbf{cheated}}}
         \put(32,87){\boldmath\color{white}\large{\textbf{$|m^{(\rho)}-\rho|$}}}
         \put(24.1,55.65){\boldmath\color{white}$\bullet$}
         \put(24.1,55.65){\boldmath\color{black}$\circ$}
     \end{overpic}
  \\
  \end{tabular}
 \caption{Top-down view  on the Galactic plane, showing the uncertainty estimate ($\sigma^{(\rho)}$, left panels) and the absolute error ($|m^{(\rho)}-\rho$|, right panels) for our reconstruction $m^{(\rho)}$ in units of $\mathrm{cm}^{-3}$. The input density $\rho$ has fluctuations of medium strength and is sampled by 5000 pulsars with $5\%$ distance uncertainty.
The top row shows the scenario with unknown power spectrum and Galactic profile, the bottom row shows the scenario with known power spectrum and profile. We note that the left and right panels have different color bars.}
 \label{fig:uncertainty}
\end{figure*}

Here, we discuss the $1\sigma$ uncertainty map of the reconstruction of the Galaxy model with medium fluctuations using the data set with 5000 pulsars and $5\%$ distance uncertainty. We compare the uncertainty map for unknown profile and power spectrum and the uncertainty map for known profile and power spectrum (see Appendix~\ref{sec:cheated}) with the corresponding absolute errors.
The density maps can be seen in Fig.~\ref{fig:cheated}.
The uncertainty estimates $\sigma^{(\rho)}$ (see Eq.~\eqref{eq:uncertainty}) are shown in Fig.~\ref{fig:uncertainty}. These uncertainties are underestimated as they are calculated from the curvature of the negative log-posterior around its minimum (see Eq.~\eqref{eq:inverse_Hessian}), not from the full distribution. By visual comparison with the absolute error\footnote{ 
To calculate the absolute error the original Galaxy model is downsampled to the resolution the algorithm uses.
}, $|m^{(\rho)}-\rho|$, we estimate that the uncertainty estimates are underestimated by a factor of roughly 3. However, their morphology seems to be reliable.

\section{Uncertainties of the vertical fall-off}
\label{sec:z_profile_uncertainity}

In principle the posterior variance of $\alpha$ and $\beta$ is the diagonal of the operator $D_{(\xi)}$ (see Eq.~\eqref{eq:profile_Hamiltonian}). However this diagonal is too large, since $\alpha$ and $\beta$ are completely degenerate with respect to a constant shift ($\alpha+c$ and $\beta-c$ yield the same profile as $\alpha$ and $\beta$). This degeneracy yields a large point variance, which is not instructive for quantifying the uncertainty of the vertical fall-off. Therefore, we project out the eigenvector corresponding to the constant shift before calculating the diagonal of $D_{(\xi)}$. This corrected diagonal is the squared $1\sigma$ uncertainty that we plot in Fig.~\ref{fig:compilation_zprof}.

\end{appendix}

\bibliographystyle{aa}
\bibliography{NE_map}

\end{document}